\documentclass[twocolumn,aps,prb,showkeys,superscriptaddress,letterpaper]{revtex4-2}
\usepackage[utf8]{inputenc}
\usepackage{amsmath}
\usepackage{amssymb}
\usepackage{graphicx}
\usepackage{bbold}
\usepackage{ulem}
\usepackage[caption=false]{subfig}
\graphicspath{{plots/}}
\usepackage[usenames,dvipsnames]{xcolor}
\usepackage{bm}
\renewcommand{\mathbf}{\bm}

\usepackage{comment}
\usepackage[unicode=true,
 bookmarks=true,bookmarksnumbered=false,bookmarksopen=false,
 breaklinks=false,pdfborder={0 0 0},pdfborderstyle={},backref=false,colorlinks=true,]
 {hyperref}
\hypersetup{linkcolor=blue,citecolor=blue,urlcolor=blue}

\begin{document}

\title{Spin-polarized transport and quantum phase transitions in one-dimensional superconductor-ferromagnetic insulator heterostructures}

\author{Javier Feijóo}

\affiliation{Instituto de Física La Plata - CONICET, Diag 113 y 64 (1900) La Plata, Argentina}
\affiliation{Departamento de Física, Universidad Nacional de La Plata, cc 67, 1900 La Plata, Argentina.}

\author{Aníbal Iucci}

\affiliation{Instituto de Física La Plata - CONICET, Diag 113 y 64 (1900) La Plata, Argentina}
\affiliation{Departamento de Física, Universidad Nacional de La Plata, cc 67, 1900 La Plata, Argentina.}

\author{Alejandro M. Lobos}

\affiliation{Facultad de Ciencias Exactas y Naturales, Universidad Nacional de Cuyo and CONICET, 5500 Mendoza, Argentina}
\affiliation{Instituto Interdisciplinario de Ciencias Básicas (CONICET-UNCuyo)}

\begin{abstract}
We theoretically propose a one-dimensional electronic nanodevice inspired in recently fabricated semiconductor-superconductor-ferromagnetic insulator (SE-SC-FMI) hybrid heterostructures, and  investigate its subgap transport properties. While previous related studies have primarily focused on the potential for generating topological superconductors hosting Majorana fermions, we propose an alternative application for these devices: to generate tunable spin-polarized Andreev 
 bound states (ABS) with potential uses in the 
 design of spintronic devices. The proposed setup allows to controllably explore and detect the subgap ABS and to identify the associated spin- and parity-changing transitions in tunel transport experiments. Our study highlights two key differences from existing devices: first, the length of the FMI layer must be shorter than that of the SE-SC heterostructure, introducing an inhomogeneous Zeeman interaction with significant effects on the induced ABS. Second, we focus on semiconductor nanowires with minimal or no Rashba spin-orbit interaction, allowing for the induction of spin-polarized ABS and high-spin quantum ground states. We show that the device can be tuned across spin- and fermion parity-changing transitions by adjusting the FMI layer length and/or by applying a global back gate voltage, with zero-energy crossings of subgap ABS as signatures of these transitions. Our findings suggest that these effects are experimentally accessible and offer a robust platform for studying and controlling spin-polarized ABS and quantum phase transitions in hybrid nanowires.
\end{abstract}

\maketitle

\section{Introduction}

Hybrid heterostructures combining properties of different materials have emerged as new platforms to investigate quantum phases and exotic phenomena otherwise nonexistent in nature. A prominent example are superconductor-ferromagnet (SC-FM) heterostructures \cite{Huebler12_ABS_at_spin_active_interfaces, Wolf14_Spin_polarized_transport_at_Al_EuS_interfaces, Buzdin05_Review_SC_FM_heterostructures, Bergeret05_Review_SC_FM_heterostructures,  Eschrig18_Review_ABS_in_SFS_junctions,
Bergeret18_Nonequilibrium_effects_in_SC_FM_structures, Heikkila19_Review_FMI_SC_structures}, which combine  
two antagonistic states of matter: ferromagnets, where the spins of electrons tend to be aligned, and superconductors, where their microscopic constituents, the Cooper pairs, are made of electrons with anti-aligned spins. When these materials are forced to coexist at an interface, a rich phenomenology emerges, such as the spatial oscillation of superconducting correlations \cite{Fulde64_FFLO, Larkin64_FFLO}, the emergence of $\pi-$junction behavior in SC-FM junctions (see Ref. \cite{Buzdin05_Review_SC_FM_heterostructures} and references therein), the generation of spin-triplet superconductors \cite{Diesch18_Equal_spin_triplet_superconductivity_at_EuS_Al_interfaces,Minutillo2021}, and the generation of spin-split superconductors (see Ref. \cite{Bergeret18_Nonequilibrium_effects_in_SC_FM_structures} and references therein). From the technological point of view, SC-FM hybrids represent a potential breakthrough for low-dissipation electronics and the generation of spin-polarized transport \cite{Huebler12_ABS_at_spin_active_interfaces, Wolf14_Spin_polarized_transport_at_Al_EuS_interfaces}

Another example of advanced hybrid materials arises in the context of Majorana quantum wires, where semiconductor-superconductor (SE-SC)  heterostructures have been fabricated to furnish a semiconductor nanowire with superconducting correlations via the proximity effect  (see Refs. \cite{Mourik12_Signatures_of_MF, Das12_Evidence_of_MFs, Prada2020} and references therein). 

According to theoretical works \cite{Lutchyn10_MF_and_Topological_transition_in_SM_SC_Heterostructures, Oreg10_Helical_liquids_and_MF_in_QW}, a topological superconductor ground state with Majorana zero-modes localized at the ends of the device could be induced as a result of the combination of three main ingredients: a) strong Rashba spin-orbit coupling, such as that existing in InAs or GaAs semiconductor nanowires, b) proximity-induced superconductivity provided by the interface with a conventional $s-$wave superconductor (e.g., Al), and c) an externally applied magnetic field in the direction perpendicular to the spin-orbit axis. 

\begin{figure}
    \includegraphics[width=0.98\columnwidth]{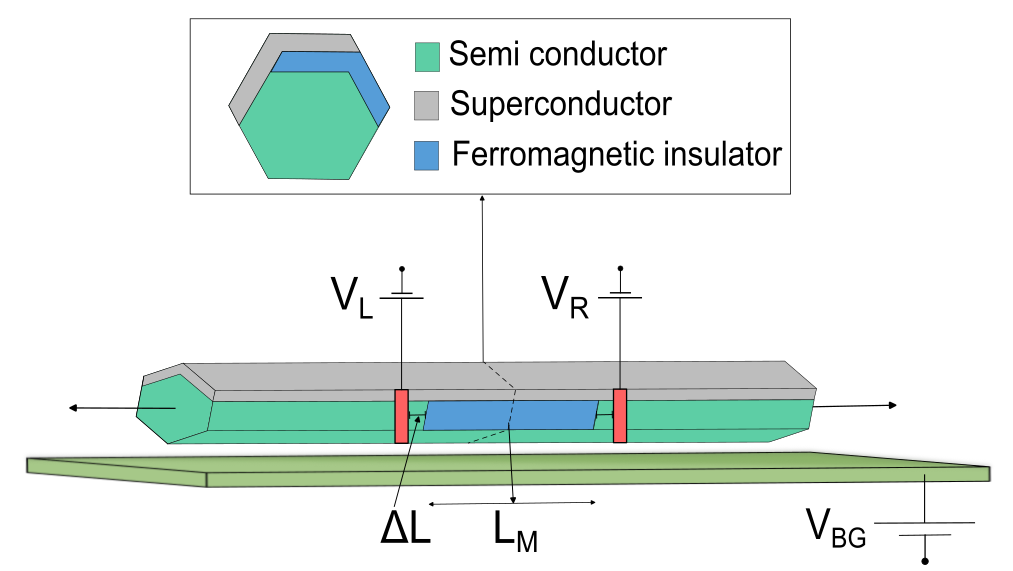}
    \caption{(color online) Schematic representation of the SC-SE-FMI heterostructure. The contact leads (denoted in red) are coupled to
 the SE nanowire, at each side of the FMI region at a distance $\Delta L$, and are connected to bias voltages $V_L$ and $V_R$. The backgate voltage $V_\text{BG}$ controls the chemical potential of the system.}
    \label{fig:model scheme}
\end{figure}

In recent years, a novel type of one-dimensional heterostructure combining all the aforementioned materials: semiconductors, superconductors, and ferromagnetic insulators (``SE-SC-FMI hybrids" for short) has been fabricated using molecular beam-epitaxy techniques \cite{Liu19_SM_FMI_SC_epitaxial_nanowires, Vaitiekenas21_ZBPs_in_FMI_SC_SM_hybrid_nanowires, Vaitiekenas22_Evidence_of_spin_polarized_ABS}. The main goal in these works was to avoid the necessity to apply an external magnetic field in Majorana nanowire devices, which technically constrains the device layout and is detrimental to the bulk superconductor. 
In these devices, the layer of FMI material (in this case, a EuS shell, see Fig. \ref{fig:model scheme}) generates a proximity-induced built-in Zeeman exchange field  on the SE nanowire, therefore mimicking the presence of an applied external field. In this way the operation of the triple-hybrid Majorana device can be greatly simplified. The experimental results of Ref. \cite{Vaitiekenas21_ZBPs_in_FMI_SC_SM_hybrid_nanowires} have shown suggestive zero-bias conductance peaks, potentially indicating the presence of Majorana zero-modes. Theoretical studies \cite{Woods21_Charge_impurity_effects_in_MNWs,Maiani2021,Liu21_Electronic_properties_of_SE_SU_FMI_hybrids,Langbehn2021,Khindanov2021,Escribano22_SE_SC_FMI_planar_1D_structures, Singh23_Conductance_spectroscopy_of_Majorana_SE_SC_FMI_heteorstructure, Feijoo23_ABS_and_QPT_in_SE_SC_FMI_hybrids} have recently focused on this novel setup.

Motivated in these new developments, in this article we theoretically study the quantum transport properties of a triple SE-SC-FMI hybrid device depicted in Fig. \ref{fig:model scheme}, which is inspired in (albeit different to) those studied in Refs. \cite{Liu19_SM_FMI_SC_epitaxial_nanowires, Vaitiekenas21_ZBPs_in_FMI_SC_SM_hybrid_nanowires,
Vaitiekenas22_Evidence_of_spin_polarized_ABS}.  In a previous work \cite{Feijoo23_ABS_and_QPT_in_SE_SC_FMI_hybrids}, we demonstrated that 
this type of device has interest \textit{beyond} Majorana physics due to its complex subgap electronic structure, which arises due to the existence of spin-polarized Andreev bound states (ABS) localized at the FMI region, and predicted the emergence of spin- and parity-changing quantum phase transitions (QPTs)  as a function of the parameter $k_FL_\text{M}$, where $L_\text{M}$ is the length of the FMI layer and $k_F$ is the Fermi wavelength of the semiconductor nanowire. Experimentally, for a given parameter $k_F$, the length of the magnetic region $L_\text{M}$ can be modified using recently-developed shadow-wall MBE techniques (e.g., \cite{Carrad20_Shadow_Wall_Epitaxy_of_SE_SU_Hybrids, Heedt21_Shadow_Wall_Epitaxy_of_SE_SU_Hybrids}).

In this article, we complement our previous study in Ref. \cite{Feijoo23_ABS_and_QPT_in_SE_SC_FMI_hybrids} by focusing on the effect of a uniform voltage gate, which enables the \textit{in situ} modification of the ABS energy within the gap. Physically, a uniform gate voltage effectively modifies the conduction-band filling and therefore the value of $k_F$. In addition, we compute the quantum transport properties through the device and demonstrate that the spectra of spin-polarized ABS can be experimentally accessed via, e.g., tunneling spectroscopy at the ends of the magnetic layer. We describe the system with an effective discrete model and implement the standard non-equilibrium Keldysh Green's function method to obtain the transport properties at $T=0$. Our results indicate that the application of an external back-gate indeed allows to modulate the charge of the entire SE nanowire and to modify \textit{in situ} the energy of the spin-polarized ABS. Moreover, the associated QPTs predicted to occur whenever a particle-hole pair of spin-polarized ABS crosses the Fermi level should also be experimentally accessible.

It is worth mentioning that our study is different from previous SE-SC-FMI hybrid studies in two crucial aspects:
\begin{enumerate}
    \item We assume that the length $L_\text{M}$ of the FMI layer is shorter than the length of the SE-SC heterostructure, effectively creating an inhomogeneous Zeeman exchange interaction profile which greatly affects the energy of the induced subgap ABS in the device. This setup, which can be implemented with shadow-wall masks in the MBE nanofabrication \cite{Carrad20_Shadow_Wall_Epitaxy_of_SE_SU_Hybrids, Heedt21_Shadow_Wall_Epitaxy_of_SE_SU_Hybrids}, offers a unique experimental platform to design quantum states and to gain control over exotic quantum groundstates. In particular the limit $k_F L_\text{M} < 1$ becomes relevant to generate isolated spin-polarized ABS, and allows to make an interesting connection with the physics of magnetic impurities in superconductors and the emergence of the ABS (known as Yu-Shiba-Rusinov states in this context, see Ref. \cite{Balatsky_2006} for a review).  
    \item We focus on SE nanowires with weak (or no) Rashba spin-orbit coupling. Although this coupling is crucial for inducing a topological phase, it prevents ABS from having a well-defined spin projection $S^z$. In our case, the spin projection of the induced ABS remains a good quantum number, enabling for spin-polarized transport and for the emergence of high-spin ground states. From an experimental point of view, the realization of such a device would require to use centrosymmetric SE nanowire, such as silicon (Si), in order to minimize the Rashba coupling and maintain $S_z$ a good quantum number. 
\end{enumerate}

The rest of the paper is organized as follows. In Sec. \ref{sec:model}, we introduce a theoretical model for a one dimensional SE-SC heterostructure with an FMI interface, including a coupling to normal leads. In Sec. \ref{sec:Transport} we present the calculation of the conductance via non-equilibrium Keldysh Green's functions. In Sec. \ref{sec:results} we present results for the local and non-local conductances, and in Sec. \ref{sec:summary}, we provide a summary and our conclusions. Finally, in the Appendix we present technical details of the theoretical methods.

\section{Theoretical model}\label{sec:model}

We theoretically describe the system with a discrete $N$-site one-dimensional model Hamiltonian representing the finite central segment containing the triple SE-SC-FMI interface (see central region in Fig. \ref{fig:model scheme}). This is an \textit{effective} Hamiltonian in which we assume a single-channel SE nanowire covered both by a bulk SC layer (e. g., an Al layer depicted in gray) and by a coexisting finite FMI layer (e. g., EuS layer depicted in blue) of length $L_\text{M}$. Then, the Hamiltonian of the central region reads
\begin{align} 
H^\text{w}_{0}&= -t\sum_{j=1,\sigma}^{N-1}\left(c_{j,\sigma}^{\dagger}c_{j+1,\sigma}+\textrm{H.c.}\right)\nonumber \\
&+\sum_{j=1,\sigma}^{N} \left[-\mu_\text{BG}\   c_{j,\sigma}^{\dagger}c_{j,\sigma}+\Delta\left(c_{j,\uparrow}^{\dagger}c_{j,\downarrow}^{\dagger}+\textrm{H.c.} \right)\right]\nonumber\\
&-h_0\sum_{j=2}^{N-1}\left(c_{j,\uparrow}^{\dagger} c_{j,\uparrow}-c_{j,\downarrow}^{\dagger} c_{j,\downarrow}\right),\label{eq:H0}
\end{align}
where \(c_{j,\sigma}^{\dagger}\) and \(c_{j,\sigma}\) represent creation and annihilation fermionic operators, respectively, at site \(j\) with spin projection \(\sigma = \{\uparrow,\downarrow\}\). The parameter \(t\) describes the effective hopping amplitude of electrons in the single-channel SE nanowire, the effective pairing parameter \(\Delta\) represents the proximity-induced superconducting pairing amplitude generated by the coupling with the SC layer, and \(h_0\) corresponds to the effective Zeeman exchange interaction induced by the FMI layer. We have excluded the Zeeman exchange $h_0$ from the sites at the ends (i.e., sites $j=1$ and $j=N$) in order to allow these sites to couple to the measuring contacts. This is a technicality that simplifies our subsequent calculations and the embedding method (see below).
Therefore, the actual length of the FMI layer is represented by $N_\text{M}=N-2$ sites (i.e., $L_\text{M}=N_\text{M} a$, where $a$ is the lattice parameter). 

The chemical potential \(\mu_\text{BG}\) is a parameter of crucial importance in our work, since it effectively incorporates the effect of a uniform back gate potential \(V_\text{BG}\) (see green surface in Fig. \ref{fig:model scheme}) acting on the entire length of the heterostructure, and allows to modulate \textit{in situ} the overall band filling of the SE nanowire. We stress that the ability to modulate the charge in these devices has been experimentally demonstrated in a recent work \cite{Vaitiekenas21_ZBPs_in_FMI_SC_SM_hybrid_nanowires}.  As stated in Ref. \cite{Feijoo23_ABS_and_QPT_in_SE_SC_FMI_hybrids}, the critical parameters controlling the position of the ABS in this type of device are the dimensionless product $k_\text{F}L_\text{M}$ and the ratio $h_0/\Delta$. However, since typically both $h_0$ and $\Delta$ are fixed by the choice of parent FMI and SC bulk materials and by the transparency of the interface, they cannot
be modified \textit{in situ} for a given device. By introducing a global background gate $V_\text{BG}$, we claim that the Fermi momentum $k_F$ can be externally controlled through the relation $\mu_\text{BG}=\hbar^2 k_F^2/2m$, therefore providing an experimentally feasible way to modify the subgap spectrum and to explore the phase diagram of the device.

In addition to the SE-SC-FMI interface, the device is assumed to be weakly coupled to measuring normal leads connected to bias voltages $V_L$ and $V_R$, introduced to study its transport properties (see Fig. \ref{fig:model scheme}). In our model, the measurement-contact leads are described by the Hamiltonian
\begin{align} \label{eq: H_tr}
H^\text{leads}&=H^\text{lead}_{L}+H^\text{lead}_{R}+H^\text{lead-w},
\end{align}
where the terms $H^\text{lead}_L$ and $H^\text{lead}_R$ describe, respectively, the left and right normal leads (see  red finger-type structures in Fig. \ref{fig:model scheme}) which we  describe as non-interacting semi-infinite chains
\begin{align}
    H^\text{lead}_\alpha =-\sum_{j=0,\sigma}^\infty \left[ \mu_\alpha d^\dagger_{\alpha,j;\sigma}d_{\alpha,j;\sigma}+\left(t^\prime d^\dagger_{\alpha,j;\sigma}d_{\alpha,j+1;\sigma}+\text{H.c.}\right)\right],\label{eq:H_leads}
\end{align}
where $\alpha=\left\{L,R\right\}$, and where $\mu_\alpha=-eV_\alpha$ are the respective chemical potentials of the leads, which can be independently controlled via bias voltages $V_\alpha$ and can be used to perform tunneling spectroscopy of the device. Finally, the term
\begin{align}
    H^\text{lead-w}&=\sum_\sigma \left[ t_L d^\dagger_{L,0;\sigma}c_{1;\sigma} + t_R d^\dagger_{R,0;\sigma}c_{N;\sigma} +\textrm{H.c.} \right],\label{eq:Hmix_leads}
\end{align}
represents the tunneling coupling, where the tunneling parameters $t_L,\ t_R$ can be independently controlled via pinch-off gates placed beneath the SE-SC-FMI structure (not shown in Fig. \ref{fig:model scheme}).

A crucial aspect in the theoretical study of this type of nanostructures is the fact that the localization length of the subgap ABS, $\xi\sim \hbar v_F/\sqrt{\Delta^2-\epsilon_\text{ABS}^2}$, where $v_F$ is the Fermi velocity of the SE nanowire and $\epsilon_\text{ABS}$ is the energy of the ABS inside the gap $\Delta$,  can greatly exceed  $L_\text{M}$ when $\left|\epsilon_\text{ABS}\right|\to \Delta$, introducing important finite-size effects. While in certain experimental devices such finite-size effects are a real consequence of quantum confinement, for simplicity here we have eliminated them to avoid further theoretical complexities. A basic strategy to minimize these effects consists in increasing the system size $L_\text{M}$ to values $L_\text{M}\gg \xi$. However, such na\"ive procedure can rapidly exceed the capability of our computational resources, and therefore other approaches are needed. In this work we use Dyson's equation to embed the finite-wire+leads system into semi-infinite one-dimensional chains located on each side of the central region (see Fig. \ref{fig:model scheme}), therefore reconstituting an infinite heterostructure where finite-size effects are absent. 
We introduce the semi-infinite 1D chains representing SE-SC heterostructures (with no FMI layer) 
\begin{align}
H^\text{w}_\text{I}&=\sum_{j=0,\sigma}^{-\infty} \left[-\mu_\text{BG}\   c_{j,\sigma}^{\dagger}c_{j,\sigma} -t\left(c_{j,\sigma}^{\dagger}c_{j+1,\sigma}+\textrm{H.c.}\right)\right],\nonumber\\
&+\Delta\sum_{j=0}^{-\infty}\left(c_{j,\uparrow}^{\dagger}c_{j,\downarrow}^{\dagger}+\textrm{H.c.} \right)\label{eq:HwL}\\
H^\text{w}_\text{II}&=\sum_{j=N+1,\sigma}^{\infty} \left[-\mu_\text{BG}\   c_{j,\sigma}^{\dagger}c_{j,\sigma} -t\left(c_{j,\sigma}^{\dagger}c_{j+1,\sigma}+\textrm{H.c.}\right)\right]\nonumber\\
&+\Delta\sum_{j=N+1}^{\infty}\left(c_{j,\uparrow}^{\dagger}c_{j,\downarrow}^{\dagger}+\textrm{H.c.} \right)\label{eq:HwR},
\end{align}
and the coupling term

\begin{align} 
H^\text{w}_\text{mix}&=-t\sum_\sigma \left(c_{0,\sigma}^{\dagger}c_{1,\sigma}+c_{N,\sigma}^{\dagger}c_{N+1,\sigma}+\textrm{H.c.}\right),\label{eq:Hmix_wire}
\end{align}
that merges the three Hamiltonians $H_0^\text{w}$, $H_\text{I}^\text{w}$, and $H_\text{II}^\text{w}$ together. For simplicity, and to avoid the introduction of further inhomogeneities, identical  values of $t,\ \mu_\text{BG}$ and $\Delta$  are used in Eqs. (\ref{eq:H0}), (\ref{eq:HwL}), and (\ref{eq:HwR}).
This embedding procedure allows us to study
the effect of an inhomogeneous Zeeman exchange field of length $L_\text{M}$ on an infinitely-long, and otherwise homogeneous, SE-SC heterostructure at a very low computational cost. In practice, we expect this model to be valid in the limit where the total length of the device $L_\text{T}$ is $L_\text{T} \gg \xi$.  

Concerning the position of the measuring leads, it is important to realize that due to the localization of the ABS wavefunctions, the contacts 
cannot be placed too far from the FMI region (i.e., at distances much larger than $\xi$), otherwise there would be no overlap between the ABS and the normal quasiparticles' wavefunctions in the contact zone, and therefore no ABS-mediated current would be detected.

We now briefly comment on the choice of parameters in our model. In our numerical calculations, we choose parameters that represent the experimental systems as close as possible. Note that the Hamiltonian Eq. (\ref{eq:H0}) must be interpreted as a coarse-grained model describing effective degrees of freedom in the SE-SC-FMI structure near the Fermi level and at low temperatures $T\ll \Delta$. Here, both the SC and FMI layers have been integrated out, and we are assuming that their effects are entirely described by effective parameters $\Delta$ and $h_0$ respectively.
In particular, the hopping $t$, representing the effective bandwidth of the SE nanowire, is an arbitrary parameter that needs to be fitted to reproduce the low-energy physics near $E_F$. In order to fix it, we note that in most physical situations the relation $\Delta \ll t$ is obeyed. Therefore, we have set the ratio $\Delta/t$ to the conservative value $\Delta/t=0.1$. Next,  we have set the relation between the induced pairing parameter $\Delta$ and the Zeeman interaction $h_0$ as $h_0/\Delta=1.5$ according to the results reported in Ref. \cite{Vaitiekenas21_ZBPs_in_FMI_SC_SM_hybrid_nanowires} for InAs-Al-EuS in hybrid nanowires. Finally, assuming that the physical superconducting coherence length induced in the NW $\xi_0=\hbar v_F/\Delta$ is of the order of $\xi_0 \approx 50$ nm \footnote{S. Vaitiek{\.{e}}nas, private comunication}, and using the expression of the Fermi velocity extracted from the tight-binding model at half-filling $k_F a=\pi/2$, i.e.,  $ v_F = t a/\hbar\pi$, we obtain the estimation for the coarse-grained lattice parameter $a\simeq 18 $ nm. With these estimated values, the length $L_\text{M}$ used in Ref. \cite{Vaitiekenas22_Evidence_of_spin_polarized_ABS}  corresponds to a system of $N_\text{M}\approx 25$ magnetic sites. We note, however, that we are \textit{not} limited to these system sizes in our numerical calculations. In addition, we recall that our \textit{total} system size is actually infinite, due to the embedding procedure (see Appendix \ref{Ap:embedding}).

We finally note that many experimental details, beyond the scope of the present work, have been neglected in the above model, such as the effect of  of electron-electron interactions, or the non-trivial simultaneous coexistence of SC-FMI layers, whose effects greatly differ depending on the stacking of the layers \cite{Vaitiekenas22_Evidence_of_spin_polarized_ABS, Escribano22_SE_SC_FMI_planar_1D_structures}. Another important non-ideality is the presence of disorder. While disorder is not expected to be a major issue in MBE-grown devices, it is still a conceptually important perturbation that needs to be addressed theoretically. In Sec. \ref{sec:results} we include diagonal on-site disorder to test the robustness of our predictions.

\section{Calculation of the transport properties}\label{sec:Transport}

To obtain the transport properties of this system, we first discuss the general solutions of the isolated SE-SC-FMI heterostructure described by Eq. (\ref{eq:H0}) using the Bogoliubov-de Gennes formalism. Then, in Appendix \ref{app:coupling_to_the_leads} we show the derivation of the analytical expressions used to compute the tunneling conductance. In Appendix \ref{app:embedding} we show details of the calculation of the dressed Green's function after coupling the central region to the measuring contacts.

\subsection{Isolated SE-SC-FMI heterostructure}\label{sec:isolated_system}

We now focus on the Hamiltonian $H_0^\text{w}$ in Eq. (\ref{eq:H0}), which describes the central finite region in Fig. \ref{fig:model scheme}, and define the following Nambu spinors
\begin{align}\label{eq:Psi}
\Psi_\uparrow & \text{=}\begin{pmatrix}
\boldsymbol{\psi}_\uparrow\\
\boldsymbol{\psi}^\dagger_\downarrow
\end{pmatrix},& 
\Psi_\downarrow & =\begin{pmatrix}
\boldsymbol{\psi}_\downarrow\\
\boldsymbol{\psi}^\dagger_\uparrow
\end{pmatrix},
\end{align}
where we have introduced the $N-$component vectors $\boldsymbol{\psi}_\sigma=\left\{c_{1,\sigma}, c_{2,\sigma},\dots,c_{N,\sigma} \right\}^T$. Note that the Nambu vectors in Eq. (\ref{eq:Psi}) have definite spin projection even though the charge is not a good quantum number. With these definitions we can compactly write the Hamiltonian $H^\text{w}_0$ as
\begin{equation} \label{eq:H0_BdG}
{H}^\text{w}_{0}=\frac{1}{2}\sum_\sigma \Psi_\sigma^{\dagger}\mathcal{H}^\text{w}_{0,\sigma}\Psi_\sigma+E_{0},
\end{equation}
where the constant term is $E_{0}=-\frac{1}{2}\sum_\sigma\text{Tr}\{\mathcal{H}^\text{w}_{0,\sigma}\}$. In terms of this basis, the matrix representation of the 
Bogoliubov-de Gennes (BdG) Hamiltonian is 
\begin{equation} \label{eq:H0sigma}
\mathcal{H}^\text{w}_{0,\sigma}=\begin{pmatrix}
\hat{T}+\sigma \hat{h} & -\sigma \hat{\Delta}\\
-\sigma \hat{\Delta}^\dagger & -(\hat{T}-\sigma \hat{h})
\end{pmatrix}.
\end{equation}
Here, \(\hat{T}\), \(\hat{h}\)  and \(\hat{\Delta}\) are $N\times N$ matrices defined as \\
$\left[\hat{T}\right]_{i,j}= -t (\delta_{i,j+1}+\delta_{i+1,j}) -\mu_\text{BG} \delta_{i,j}$, $\left[\hat{h}\right]_{i,j}=h_0\delta_{i,j}$ (for $\{i,j\} \in \{2,\dots,N-1\}$),  $\left[\hat{\Delta}\right]_{i,j}=\Delta \delta_{i,j}$
corresponding, respectively, to the tight-binding, Zeeman, and pairing potential terms in Eq. (\ref{eq:H0}).
We note that the Hamiltonian Eq. (\ref{eq:H0sigma}) satisfies the anti-unitary particle-hole symmetry 
\begin{equation} \label{eq:particle_hole _symmetry}
\mathcal{H}^\text{w}_{0,\uparrow}=-\tau_{x}\mathcal{H}^{\text{w}\dagger}_{0,\downarrow} \tau_{x},
\end{equation}
where the Pauli matrix $\tau_{x}$ acts on the Nambu space. Therefore, if $\Phi_{\nu,\sigma}$ is an eigenvector of $\mathcal{H}^\text{w}_{0,\sigma}$ with eigenvalue $\epsilon_{\nu,\sigma}$, i.e., 
\begin{align}
\mathcal{H}^\text{w}_{0,\sigma}\Phi_{\nu,\sigma}=\epsilon_{\nu,\sigma}\Phi_{\nu,\sigma},\label{eq:eigenvalueBdG}
\end{align}
the vector $\Phi_{\nu,\bar{\sigma}}=\tau_{x}\Phi^{*}_{\nu,\sigma}$ is an eigenvector of $\mathcal{H}^\text{w}_{0,\bar{\sigma}}$ with eigenvalue $-\epsilon_{\nu,\sigma}$. Explicitly, the eigenvectors $\Phi_{\nu,\sigma}$ and $\Phi_{\nu,\bar{\sigma}}$ take the form
\begin{align} \label{eq:eigenvectors}
\Phi_{\nu,\sigma}&=\begin{pmatrix}
\mathbf{u}_{\nu,\sigma}\\ 
\mathbf{v}_{\nu,\sigma}
\end{pmatrix},&\Phi_{\nu,\bar{\sigma}}&=\begin{pmatrix}
\mathbf{v}^*_{\nu,\sigma}\\ 
\mathbf{u}^*_{\nu,\sigma}
\end{pmatrix},
\end{align} 
with $N$-component vectors $\mathbf{u}_{\nu,\sigma}=\{u^\sigma_{1,\nu}, u^\sigma_{2,\nu}, \dots, u^\sigma_{N,\nu}\}^T$ and $\mathbf{v}_{\nu,\sigma}=\{v^\sigma_{1,\nu}, v^\sigma_{2,\nu}, \dots, v^\sigma_{N,\nu}\}^T$, where $u^\sigma_{j,\nu}$ and $v^\sigma_{j,\nu}$ are complex coefficients. The eigenmodes $\Phi_{\nu,\sigma}$ obey the orthogonality condition
\begin{align}
\Phi^\dagger_{\nu,\sigma}\cdot\Phi_{\mu,\sigma}&=\delta_{\nu,\mu}.
\end{align}

Finally, the destruction operators of quasi-particles (i.e., the ``Bogoliubons") can be written as
\begin{equation}\label{eq:eigenmode}
\gamma_{\nu,\sigma}=\sum_{j=1}^{N}\left[\left(u^\sigma_{j,\nu}\right)^* c_{j,\sigma}+\left(v^\sigma_{j,\nu}\right)^*c_{j, \bar{\sigma}}^\dagger \right],
\end{equation}
and from unitarity we obtain the eigenmode expansion of the original fermionic operators
\begin{align}
c_{j,\sigma}&=\sum_{\nu=1}^{2N} u^\sigma_{j,\nu} \gamma_{\nu,\sigma}, \label{eq:eigenmode_expansion_c}\\ 
c^\dagger_{j,\bar{\sigma}}&=\sum_{\nu=1}^{2N} v^\sigma_{j,\nu}\gamma_{\nu,\sigma}. \label{eq:eigenmode_expansion_c_dagger}
\end{align}

For later convenience, we define the (retarded) Green's function matrix in Nambu space corresponding to the Hamiltonian Eq. (\ref{eq:H0}) as \cite{Cuevas96_Hamiltonian_approach_to_SC_contacts}
\begin{align}
\hat{\mathbf{g}}_{i,j;\sigma}^{(0),r}(t,t')&\equiv
\begin{pmatrix}
g_{i,j;\sigma}^{(0),r}\left(t,t^\prime \right) & f_{i,j;\sigma}^{(0),r}(t,t')\\
\bar{f}_{i,j;\sigma}^{(0),r}\left(t,t^\prime \right) & \bar{g}_{i,j;\sigma}^{(0),r}\left(t,t^\prime \right)
\end{pmatrix},\nonumber\\*[1em]
&=-i\theta(t-t^\prime)\nonumber\\*[0.4em]
&\times \begin{pmatrix}
   \langle \{c_{i,\uparrow}(t);c^\dagger_{j,\uparrow}(t^\prime) \}\rangle        &      \langle \{c_{i,\uparrow}(t);c_{j,\downarrow}(t^\prime)\}\rangle \\
   \langle \{c^\dagger_{i,\downarrow}(t);c^\dagger_{j,\uparrow}(t^\prime)\} \rangle & \langle \{c^\dagger_{i,\downarrow}(t);c_{j,\downarrow}(t^\prime)\}\rangle
\end{pmatrix}.
\end{align}
Taking the Fourier transform of this expression, and using the eigenmode solutions in Eq. (\ref{eq:eigenvectors}), we obtain
\begin{align}
\hat{\mathbf{g}}_{i,j;\sigma}^{(0),r}(\omega)&=\sum_\nu \frac{1}{\omega-\epsilon_{\nu,\sigma}+i\eta} 
\begin{pmatrix}
u^{\sigma}_{i,\nu}\left(u^{\sigma}_{j,\nu}\right)^* &u^{\sigma}_{i,\nu}\left(v^{\sigma}_{j,\nu}\right)^*\\
v^{\sigma}_{i,\nu}\left(u^{\sigma}_{j,\nu}\right)^*&v^{\sigma}_{i,\nu}\left(v^{\sigma}_{j,\nu}\right)^*
\end{pmatrix}
\label{eq:g0_matrix}
\end{align}
with $\{i,j\}\in \{1,\dots,N\}$ representing sites in the central region. Note that every pair $\{i,j\}$ defines a 2$\times$2 Nambu matrix $\hat{\mathbf{g}}_{i,j;\sigma}^{(0),r}(\omega)$. The physics of the ABS is already contained here, i.e., they appear as isolated poles located at frequency $\omega=\epsilon_{\nu, \sigma}$ inside the gap.

\subsection{Transport properties and tunneling spectroscopy}\label{sec:transport_properties}

The calculation of the  Green's functions from the previous sections allows us to compute transport properties, specifically the local and non-local conductance through the device, in a computationally efficient manner. 
We start from the definition of the total charge current at the leads, as the change of the number of particles $ N_\alpha=\sum_{j,\sigma} d^\dagger_{\alpha,j,\sigma}d_{\alpha,j,\sigma}$  in the normal reservoirs with respect to time \cite{Cuevas96_Hamiltonian_approach_to_SC_contacts, meir92}
\begin{equation} \label{Eq:Number}
   I_{\alpha} =-e \left\langle \frac{dN_\alpha}{dt}\right\rangle=\frac{ie}{\hbar}\left\langle \left[H,N_{\alpha}\right]\right\rangle.
\end{equation}
The only term in the Hamiltonian that does not commute with \(N_{\alpha}\) is \(H^{\text{lead-w}}\), therefore the total current in the contacts is expressed as   
\begin{align}
I_{L}&=\frac{ie}{\hbar}t_{L}\sum_{\sigma}\left[\left\langle d_{L,0,\sigma}^{\dagger}c_{1,\sigma}\right\rangle -\left\langle c_{1,\sigma}^\dagger d_{L,0,\sigma} \right\rangle \right] \label{eq:I_L} \\
I_{R}&=\frac{ie}{\hbar}t_{R}\sum_{\sigma}\left[\left\langle d_{R,0,\sigma}^{\dagger}c_{N,\sigma}\right\rangle -\left\langle c_{N,\sigma}^\dagger d_{R,0,\sigma} \right\rangle \right].\label{eq:I_R}  
\end{align}
These quantities can be expressed in terms of the non-equilibrium Keldysh Green's functions, specifically in terms of the lesser Green's functions \cite{Haug_Jauho_book}
\begin{align}
G_{\alpha,\beta;\sigma}^{<}(t,t^\prime)&=i\left\langle f_{\beta,\sigma}^{\dagger}\left(t^\prime\right)f_{\alpha,\sigma}\left(t\right)\right\rangle,\nonumber\\
&=\int_{-\infty}^{\infty}d\omega\ e^{i\omega (t-t^\prime)} G_{\alpha,\beta;\sigma}^{<}(\omega),\label{Eq:GreenLesser}
\end{align}
where $f_{\alpha,\sigma}, f_{\beta,\sigma}$ are generic fermion operators $f=\{c,d\}$, and $\alpha,\beta$ are generic site indices. Note that $G_{\alpha,\beta;\sigma}^{<}(\omega)$  in Eq. (\ref{Eq:GreenLesser}) is the first element of the fully dressed (i.e., after coupling the leads and the semi-infinite chains) Nambu matrix Green's function. After substituting Eq. (\ref{Eq:GreenLesser}) into Eqs. (\ref{eq:I_L}) and (\ref{eq:I_R}), the currents  in terms of the Fourier-transformed lesser functions can be expressed as follows \cite{mahan}:
\begin{align}
I_{L}&=\frac{e}{h}t_{L}\sum_\sigma\int_{-\infty}^{\infty}d\omega\ \left[\hat{\mathbf{G}}_{L,1;\sigma}^{<}(\omega)-\hat{\mathbf{G}}_{1,L;\sigma}^{<}(\omega)\right]_{1,1},\label{eq:current_G_lesserL} \\
&=-\frac{e}{h}t^2_{L}\sum_\sigma\int_{-\infty}^{\infty}d\omega\ \left[\hat{\mathbf{g}}_{L,L;\sigma}^{(0),<}(\omega)\hat{\mathbf{G}}_{1,1;\sigma}^{>}(\omega)\right.\nonumber\\
&\quad \quad \left.-\hat{\mathbf{g}}_{L,L;\sigma}^{(0),>}(\omega)\hat{\mathbf{G}}_{1,1;\sigma}^{<}(\omega)\right]_{1,1},\label{eq:current_G_lesserL2}
\end{align}
and
\begin{align}
I_{R}&=\frac{e}{h}t_{R}\sum_\sigma \int_{-\infty}^{\infty}d\omega \left[\hat{\mathbf{G}}_{R,N;\sigma}^{<}(\omega)-\hat{\mathbf{G}}_{N,R;\sigma}^{<}(\omega)\right]_{1,1}, 
\label{eq:current_G_lesserR}\\
&=-\frac{e}{h}t^2_{R}\sum_\sigma\int_{-\infty}^{\infty}d\omega\ \left[\hat{\mathbf{g}}_{R,R;\sigma}^{(0),<}(\omega)\hat{\mathbf{G}}_{N,N;\sigma}^{>}(\omega)\right. \nonumber \\
&\quad \quad \left. -
\hat{\mathbf{g}}_{R,R;\sigma}^{(0),>}(\omega)\hat{\mathbf{G}}_{N,N;\sigma}^{<}(\omega)\right]_{1,1},
\end{align}
where the index $\left[\dots\right]_{a,b}$ denotes the element of the Nambu structure of the  Green's matrix Eq. (\ref{eq:G_fully_dressed}), and where we have used the Keldysh-Dyson's equations to  conveniently express the current in terms of local Green's functions of the central region \cite{meir92}. In addition, we have introduced the bare lesser and greater Green's functions of the contacts

\begin{align}
\hat{\mathbf{g}}_{\alpha,\alpha}^{(0),<}(\omega)&= 2\pi i\rho^{(0)}_{\alpha} \begin{pmatrix}
n_{\alpha} & 0\\
0 & \bar{n}_{\alpha}(\omega)
\end{pmatrix}\label{eq:lesser_density}, \\
\hat{\mathbf{g}}_{\alpha,\alpha}^{(0),>}(\omega)&= -2\pi i\rho^{(0)}_{\alpha}\begin{pmatrix}
1-n_{\alpha}(\omega) & 0\\
0 & 1-\bar{n}_{\alpha}(\omega)
\end{pmatrix},\label{eq:greater_density}
\end{align}
where $n_{\alpha}(\omega)=n_{F}(\omega+\mu_{\alpha})$ is the Fermi distribution function at the contact $\alpha$, and  $\bar{n}_{\alpha}(\omega)=1-n_{F}(\omega+\mu_{\alpha})$. Finally, the chemical potential at the contacts $\mu_\alpha = - e V_\alpha$ can be controlled with the  bias voltages $V_\alpha$ (see Fig. \ref{fig:model scheme}). Using these expressions and the relation between Green's functions $\mathbf{G}^>-\mathbf{G}^<=\mathbf{G}^r-\mathbf{G}^a$ \cite{Haug_Jauho_book}, we obtain the expressions for the currents:
\begin{widetext}
\begin{align}
I_{L}&=-\frac{ie \gamma_L}{\pi \hbar}\sum_\sigma\int_{-\infty}^{\infty}d\omega \left[n_{L}(\omega)\left(\hat{\mathbf{G}}_{1,1;\sigma}^r(\omega)-\hat{\mathbf{G}}_{1,1;\sigma}^a(\omega)\right)+\hat{\mathbf{G}}_{1,1;\sigma}^{<}(\omega)\right]_{1,1}\label{eq:I_L_final},\\
I_{R}&=-\frac{ie \gamma_R}{\hbar \pi}\sum_\sigma\int_{-\infty}^{\infty}d\omega\left[n_{R}(\omega)\left(\hat{\mathbf{G}}_{N,N;\sigma}^r(\omega)-\hat{\mathbf{G}}_{N,N;\sigma}^a(\omega)\right)+\hat{\mathbf{G}}_{N,N;\sigma}^{<}(\omega)\right]_{1,1}.\label{eq:I_R_final}
\end{align} 
\end{widetext}
By definition, the conductance is the rate of change of the current with respect to the voltage applied at the leads, and it is defined as $\mathsf{G}_{\alpha \beta}=\frac{dI_{\alpha}}{dV_{\beta}}=-e\frac{dI_{\alpha}}{d\mu_{\beta}}$. The local conductance refers to the measurement at the same site where the chemical potential is varied (i.e. $\alpha=\beta$ ), while the non-local conductance refers to the measurement at a different site (i.e. $\alpha\neq \beta$). In Eqs. (\ref{eq:I_L_final}) and (\ref{eq:I_R_final}), the chemical potential $\mu_\beta$ is implicit in the Fermi distribution functions $n_F(\omega-eV_\beta)$ and $G_{j,j;\sigma}^{<}(\omega)$. Following standard derivations outlined in  Appendix \ref{Ap:Conductance}, we obtain the expressions:

\begin{figure*}{}
        \centering
            \includegraphics[width=0.3\textwidth]{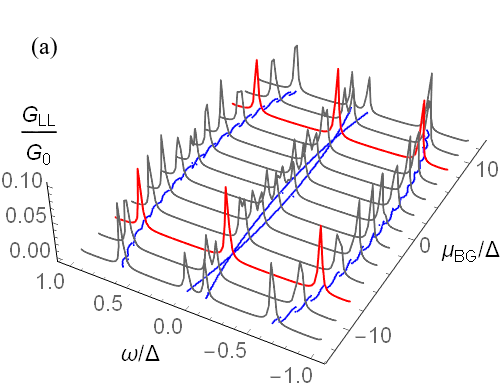} \quad
            \includegraphics[width=0.3\textwidth]{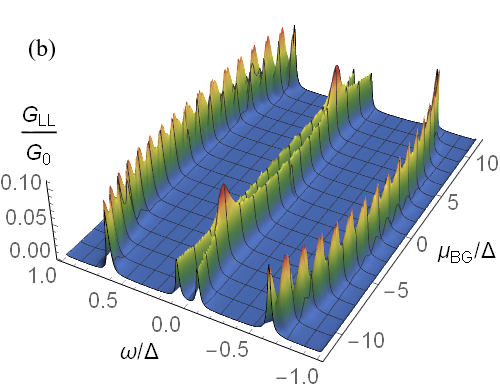} \quad
            \includegraphics[width=0.3\textwidth]{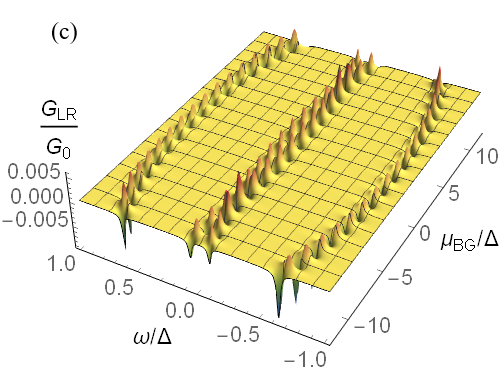}
        \caption{(color online) Local and nonlocal conductances (in units of $G_0=4e^2/h$) versus frequency $\omega$ and background chemical potential $\mu_\text{BG}$. (a) Spectral density (blue curves at basal plane) and local conductance maxima, with quantum phase transitions occurring at $\mu_\text{BG}$ values where Andreev bound states (ABS) cross $E_F$ (red curves). (b,c) High-resolution plots of local and non-local conductance, respectively, showing clearer zero-energy crossings ($\omega = 0$) of ABS in the local conductance.}
        \label{fig:conductance vs energy 2}
\end{figure*}

\begin{widetext}
\begin{align}
\mathsf{G}_{LL}(\omega,\mu)&=\frac{4e^{2}}{h} \gamma_L\sum_\sigma \left. \left\{-\text{Im}\left[G_{1,1;\sigma}^{r}(\omega,\mu)\right]+\gamma_{L}\left(-\left|G_{1,1;\sigma}^{r}(\omega,\mu)\right|^{2} + \left|F_{1,1;\sigma}^{r}(\omega,\mu)\right|^{2}\right)\right\}\right|_{\omega=-\mu_L}\label{eq:conductance_GLL},\\
\mathsf{G}_{RR}(\omega,\mu)&=\frac{4e^{2}}{h} \gamma_R\sum_\sigma \left. \left\{-\text{Im}\left[G_{N,N;\sigma}^{r}(\omega,\mu)\right]+\gamma_{R}\left(-\left|G_{N,N;\sigma}^{r}(\omega,\mu)\right|^{2} + \left|G_{N,N;\sigma}^{r}(\omega,\mu)\right|^{2} \right)\right\} \right|_{\omega=-\mu_R} \label{eq:conductance_GRR},\\
\mathsf{G}_{LR}(\omega,\mu)&=\frac{4e^{2}}{h} \gamma_L \gamma_R\sum_\sigma \left. \left(- \left|G_{1,N;\sigma}^{r}(V_R,\mu)\right|^{2} + \left|F_{1,N;\sigma}^{r}(V_R,\mu)\right|^{2} \right)\right|_{\omega=-\mu_R}  \label{eq:conductance_GLR},\\
\mathsf{G}_{RL}(\omega,\mu)&=\frac{4e^{2}}{h} \gamma_R\gamma_{L}\sum_\sigma \left|   \left(-\left|G_{N,1;\sigma}^{r}(\omega,\mu)\right|^{2} + \left|F_{N,1;\sigma}^{r}(\omega,\mu)\right|^{2} \right) \right|_{\omega=-\mu_L}, \label{eq:conductance_GRL}
\end{align}
\end{widetext}
where we used the identity  $\hat{\mathbf{G}}^{a}_{i,j;\sigma}(\omega)=\hat{\mathbf{G}}^{r}_{i,j;\sigma}(\omega)^{\dagger}$, so that the conductance is expressed solely in terms of retarded Green's functions. These formulas express the conductance in terms of the bias potentials $V_L$ and $V_R$, and the background chemical potential $\mu_\text{BG}$. This formalism provides a convenient and comprehensive framework for the numerical calculation of conductance in the system, allowing for a detailed exploration of the system's electronic transport properties in terms of parameters that can be controlled experimentally via electrical gates.

\section{Results}\label{sec:results}

We now show the results for the local \((\mathsf{G}_{LL}, \mathsf{G}_{RR})\) and non-local \((\mathsf{G}_{LR}, \mathsf{G}_{RL})\) conductances as functions of the bias potentials (i.e., $V_L$ and $V_R$) and chemical potential \(\mu_\text{BG}\) which, as stated previously, is assumed to be controlled via a back gate potential $V_\text{BG}$. Our main goal is to identify experimental signatures of spin-polarized ABS within the gap and the occurrence of the quantum phase transitions (QPTs) described above.

As is well-known \cite{Tinkham_Introduction_to_superconductivity}, local and non-local tunneling transport experiments at NS interfaces are suitable tools to explore the electronic structure of quantum devices. In our case, the electronic structure of the device originally arises from the poles of the Green's function matrix Eq. (\ref{eq:g0_matrix}) (i.e., the eigenvalues of the Hamiltonian of the isolated system $\mathcal{H}^\text{w}_{0,\sigma}$) which is dressed by the coupling to the normal leads and the coupling to the semi-infinite SE-SC wires via the Dyson's Eqs. (\ref{eq:Dyson_lead}) and (\ref{eq:Dyson_embedding}). In particular for the subgap region, tunneling transport is an experimental tool that has been successfully used to detect the presence of ABS in recent experiments \cite{Vaitiekenas21_ZBPs_in_FMI_SC_SM_hybrid_nanowires, Vaitiekenas22_Evidence_of_spin_polarized_ABS}. As discussed in these references and in seminal papers (see e.g., Ref. \cite{Sakurai70}), the QPTs occur due to an ABS crossing the Fermi energy, and are associated to changes in fermion parity and total spin $S^z$ in the groundstate of the device. 

To explicitly show the connection between the tunneling conductace and the subgap spectrum of the quantum device, in the three-dimensional plot of Fig. \ref{fig:conductance vs energy 2}\textcolor{blue}{(a)} we show the local conductance $\mathsf{G}_{LL}(\omega)$ as a function of frequency $\omega$ in the subgap region (see solid gray lines), computed for different values of $\mu_\text{BG}$ and for the particular values $N_\text{M}=\{30,38,56,82\}$,  assuming $T=0$, $\gamma_L=\gamma_R=0.1\ t$,  and the experimental ratio $h_0/\Delta=1.5$. In addition, in the basal plane of that figure we have plotted the evolution of the subgap spectrum of $\mathcal{H}^\text{w}_{0,\sigma}$ (i.e., the ABS levels) as a function of $\mu_\text{BG}$ (solid blue lines). Due to  the particle-hole symmetry of $\mathcal{H}_{0,\sigma}^\text{w}$ Eq. (\ref{eq:particle_hole _symmetry}), the subgap spectrum  shown in the basal plane of Fig. (\ref{fig:conductance vs energy 2}) reflects the symmetry $\epsilon_{\nu,\uparrow}=-\epsilon_{\nu,\downarrow}$  with respect to the line $\omega=0$.

\begin{figure*}{}
        \hspace{-2 mm}\textbf{$N_M=30$}\hspace{28 mm}\textbf{$N_M=38$}\hspace{28 mm}\textbf{$N_M=56$}\hspace{26 mm}\textbf{$N_M=82$}\hspace{10 mm}\par\medskip    
    \centering
        \includegraphics[width=0.245\textwidth]{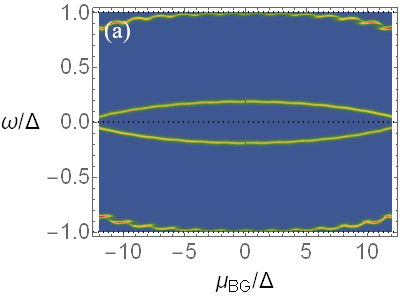} 
        \includegraphics[width=0.22\textwidth]{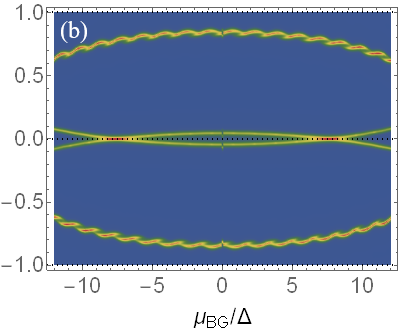}
        \includegraphics[width=0.22\textwidth]{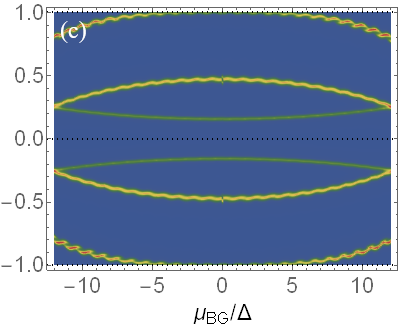} 
        \includegraphics[width=0.22\textwidth]{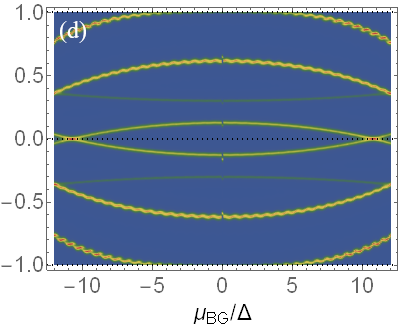} \ 
        \includegraphics[width=0.04\textwidth,height=3.2cm,trim={0 -1.2cm 0 0}]{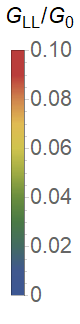}
        \includegraphics[width=0.245\textwidth]{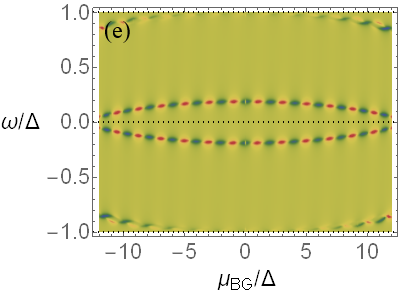}   
        \includegraphics[width=0.22\textwidth]{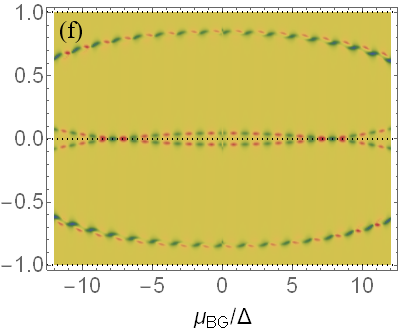}
        \includegraphics[width=0.22\textwidth]{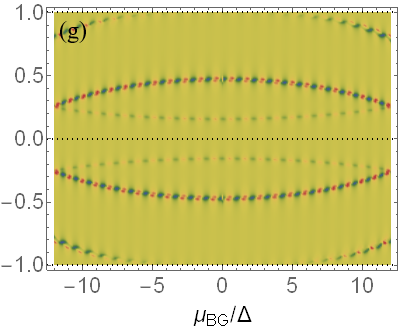}
        \includegraphics[width=0.22\textwidth]{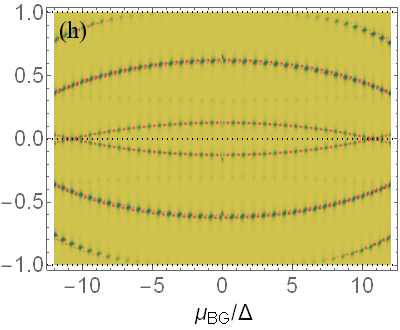} \
        \includegraphics[width=0.05\textwidth,height=3.2cm,trim={0 -1.2cm 0 0}]{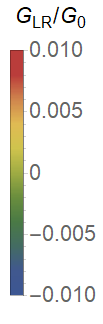}
        \includegraphics[width=0.23\textwidth]{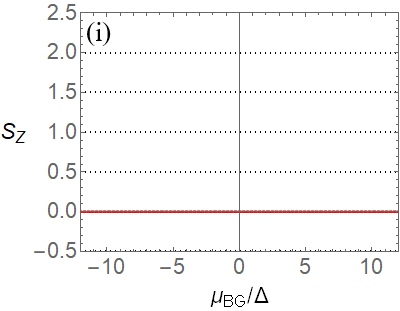}\; \ 
        \includegraphics[width=0.21\textwidth]{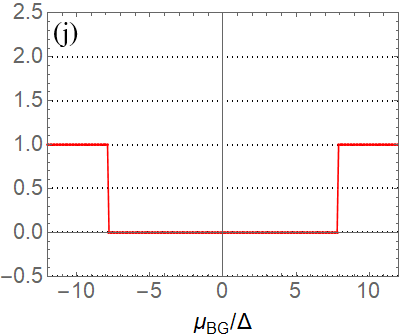}\; \ 
        \includegraphics[width=0.21\textwidth]{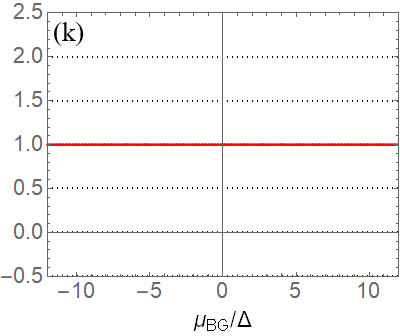}\; \ 
        \includegraphics[width=0.21\textwidth]{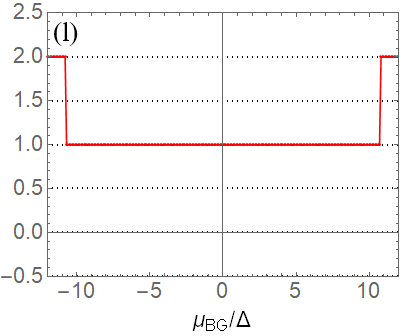}\hspace{8 mm}
        \caption{(color online)  Colormaps of (a-d) local and (e-h) nonlocal conductance (units of $G_0$) versus chemical potential $\mu_\text{BG}/\Delta$ and frequency $\omega/\Delta$, calculated for $h_0/\Delta=1.5$. (i-l) Corresponding ground state total spin $S_z$. Each column shows different ferromagnetic insulator (FMI) lengths $N_M$. Increasing $N_M$ generates spin-polarized ABS pairs from gap edges (yellow arrows), with zero-energy crossings (white arrows) appearing at $N_M=38$ ($S_z=1$) and $N_M=82$ ($S_z=2$). (i) $N_M=30$: No crossings ($S_z=0$); (j) $N_M=38$: Crossings at $\mu_\text{BG}\approx\pm8\Delta$; (k) $N_M=56$: Universal crossings (orange arrows) and new ABS group; (l) $N_M=82$: Second crossing ($S_z=2$).}  
        \label{fig:conductance for h=1.5 4L}
\end{figure*}

Note that the maxima of the resonances in $\mathsf{G}_{LL}\left(-eV_L\right)$ occur precisely for $-eV^\text{max}_L=\epsilon_{\nu,\sigma}$, as expected. When the chemical potential $\mu_\text{BG}$ is chosen such that one of the ABS levels is tuned through the $\omega=0$ line (Fermi energy), a QPT is expected to occur in the system (see solid red lines). In Fig.  \ref{fig:conductance vs energy 2}\textcolor{blue}{(b)} we show a finer mesh computation of $\mathsf{G}_{LL}(\omega)$ where the maxima of the conductance display oscillations related to the presence of nearly-degenerate ABS states with even and odd spatial inversion symmetry, due to the inversion symmetry of the Hamiltonian of the system \cite{Feijoo23_ABS_and_QPT_in_SE_SC_FMI_hybrids}. Finally, in  Fig.  \ref{fig:conductance vs energy 2}\textcolor{blue}{(c)} we show a similar calculation for the non-local conductance $\mathsf{G}_{LR}(\omega)$, which show similar albeit attenuated effects due to the exponential decay of the ABS wavefunctions.
Therefore, our results in Fig. \ref{fig:conductance vs energy 2} show that quantum transport measurement on charge-modulated devices, as the one under study, is a suitable tool to detect the occurrence of spin-polarized ABS. An important requirement for a successful detection is that the measuring contact should be placed near the FMI region, to generate enough overlap between the spin-polarized ABS and the normal quasiparticles' wavefunctions in the contacts.

Interestingly, we note that additional important information can be extracted from the transport data, in particular, from the derivative of the curve $V^\text{max}_L(\mu_\text{BG})$ with respect to $\mu_\text{BG}$. To fix ideas, let us focus on the Hamiltonian of the isolated hybrid $H_{0}^\text{w}$ in Eq. (\ref{eq:H0}). For a generic normalized  eigenstate $|\Phi_{\nu,\sigma}\rangle$, we have $\langle \Phi_{\nu,\sigma}|H_0^\text{w}|\Phi_{\nu,\sigma}\rangle=\epsilon_{\nu,\sigma}$. Applying the Hellmann-Feymann theorem \cite{feynman39} (with the difference that here we are interested in single-particle states, and not on the many-particle ground state), we have
\begin{align}
\frac{\partial \epsilon_{\nu,\sigma}}{\partial \mu_\text{BG}}&=\langle \Phi_{\nu,\sigma}|\frac{\partial H_0^\text{w}}{\partial \mu_\text{BG}}|\Phi_{\nu,\sigma}\rangle= -\langle \Phi_{\nu,\sigma}|N_0^\text{w}|\Phi_{\nu,\sigma}\rangle,\label{eq:hellmann_feynman}
\end{align}
where $N_0^\text{w}=\sum_{j=1,\sigma}^{N}   c_{j,\sigma}^{\dagger}c_{j,\sigma}$ is the total charge operator in the isolated SE-SC-FMI nanowire. Therefore, from the above arguments, we conclude that the experimental quantity $\partial V^\text{max}_{L}/\partial \mu_\text{BG}$ allows to obtain the particular \textit{average} nature (i.e., particle-like or hole-like) of the Bogoliubov-de Gennes eigenstate $|\Phi_{\nu,\sigma}\rangle$. While the above derivation has been particularized for the finite and idealized Hamiltonian $H_{0}^\text{w}$, \textit{ note that this is a generic feature of all eigenstates}. In the case of the devices discussed here, this information can be useful to detect the particle-hole symmetry point of the device as a function of $\mu_\text{BG}$, and for the particular case of the present discussion, to gain insight into the different nature of the phases before and after a QPT.

We now study the effect of varying the parameter $N_\text{M}$ (effectively representing the length of the FMI layer $L_\text{M}$) in the behaviour of the device. In Fig. \ref{fig:conductance for h=1.5 4L} we show colormaps of the local and non-local conductances (first and second rows, respectively) as functions of both the bias voltage $V_\alpha$ (represented by $\omega$) and the chemical potential $\mu_\text{BG}$.  For each column in Fig. \ref{fig:conductance for h=1.5 4L} we have used a different value of $N_\text{M}$, starting from $N_\text{M}=30$ and ending in $N_\text{M}=82$. Beginning from $N_\text{M}=30$, the curves defined by the evolution of  $V^\text{max}_L$ as function of $\mu_\text{BG}$ show the presence of 2 pairs of spin-polarized ABS, none of which cross zero, indicating that the device is in a single quantum phase characterized by $S_z=0$ (see the bottom row in Fig. \ref{fig:conductance for h=1.5 4L}), In Fig. \ref{fig:conductance for h=1.5 4L}  we have also indicated the specific spin polarization of each ABS branch. In addition, the evolution of the derivative $\partial V_L^\text{max}/\partial \mu_\text{BG}$ indicates that the two ABS in the negative frequency region $\omega <0$  \textit{have the same qualitative behavior}, changing from particle-like for $\mu_\text{BG}<0$ to hole-like for $\mu_\text{BG}>0$, as expected for Bogoliubons formed in a conduction band that is less or more than half-filled, respectively. Note in addition that the point $\mu_\text{BG}=0$, where $\partial V_L^\text{max}/\partial \mu_\text{BG}=0$, corresponds to the particle-hole symmetric point in the model, as expected. In the region $\omega>0$, the other two ABS curves correspond to the particle-hole partners of the lower ABS, which behave in the opposite way. 

The presence of only 2 ABS pairs in our calculated spectrum points to a physical situation which is qualitatively similar to that found in Ref. \cite{Vaitiekenas22_Evidence_of_spin_polarized_ABS}, where only 1  pair  of ABS (and not many) has been found experimentally. Despite the differences, this fact indicates that the parameter estimation made in Sec. \ref{sec:model} is reasonable.

\begin{figure*}{}
    \centering
        \includegraphics[width=0.245\textwidth]{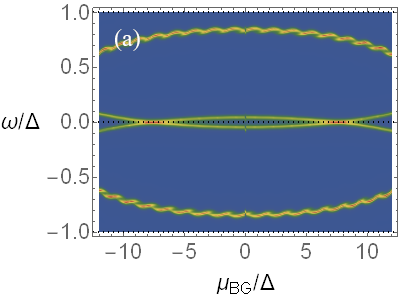} 
        \includegraphics[width=0.22\textwidth]{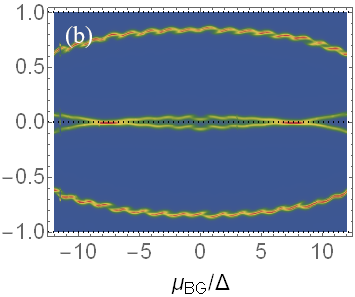}
        \includegraphics[width=0.22\textwidth]{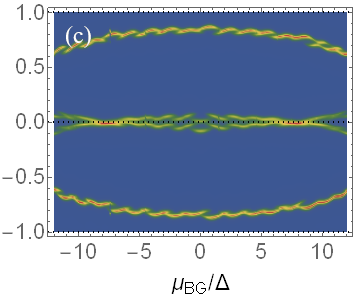} 
        \includegraphics[width=0.22\textwidth]{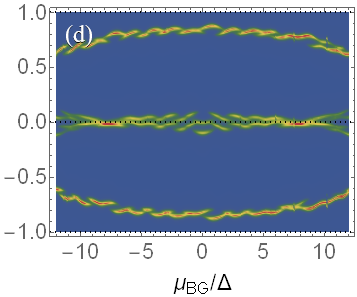} \ 
        \includegraphics[width=0.04\textwidth,height=3.2cm,trim={0 -1.2cm 0 0}]{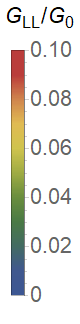}

        \caption{(color online)  Colormaps of the local conductance (in units of $G_0$) in terms of the chemical potential $\mu_\text{BG}/\Delta$ and $\omega/\Delta$, calculated for  $h_0/\Delta=1.5$ and $N_M=38$. Each column (a-d) represents different disorder strength $v_0 =  \left\{ 0,0.2,0.4,0.6\right\}$ respectively, corresponding to a weak disorder regime.}  
        \label{fig:disorder}
\end{figure*}

By increasing the length of the FMI layer to $N_\text{M}=38$, we see the appearance of two quasidegenerate spin-polarized ABS crossing zero energy at $\mu_\text{BG}\approx\pm 8 \Delta $, indicating the occurrence of QPTs at these points.  While the region $\left|\mu_\text{BG}/\Delta\right|> 8$ is characterized by $S^z=1$, the region $\left|\mu_\text{BG}/\Delta\right|< 8$ corresponds to $S^z=0$. The new subgap spectrum can be interpreted in terms of the previous one for $N_\text{M}=30$, in which the two lower (upper) ABS curves have been shifted upwards (downwards) by the effect of a larger Zeeman term due to the longer magnetic region. Increasing $N_\text{M}$ even further to $N_\text{M}=56$, we see that the region with $S^z=0$ has disappeared for the range of values $\mu_\text{BG}$ shown. Note that the first ABS curve has completely crossed the Fermi energy, and from the results of Ref. \cite{Feijoo23_ABS_and_QPT_in_SE_SC_FMI_hybrids}, we know that the position of the ABS must saturate at a value $\omega/\Delta = |\Delta-h_0|/\Delta$ in the limit $k_FL_M \to \infty$. This behavior is consistent with the results shown in Fig. \ref{fig:conductance for h=1.5 4L}, where the position of the ABS curve seem to saturate (with much lower intensity) to the value $\pm|\Delta-h_0|/\Delta= \pm 0.5$. A second ABS curve moves toward zero, while a third curve emerges from the edge of the gap. For this value of $N_\text{M}$, in both the $\omega<0$ and $\omega>0$ regions of the Figure we observe a coexistence of ABS with different curvature and therefore,  by virtue of Eq. (\ref{eq:hellmann_feynman}), of different nature (i.e., some are particle-like, other hole-like). We speculate that this coexistence is a model-independent feature of quantum phases with $S^z >0$ in this type of device.  Finally, in the last column in Fig. \ref{fig:conductance for h=1.5 4L} we show the subgap conductance corresponding to a device with a FMI layer of length $N_\text{M}=82$. In that case we observe a new QPT corresponding to a second pair of ABS crossing zero energy at $\mu_\text{BG}/\Delta \approx \pm 11$. Consequently, a ground state characterized by $S_z=2$ emerges in the region $\left|\mu_\text{BG}/\Delta\right|>11$, and by $S_z=1$ in the region $\left|\mu_\text{BG}/\Delta\right|<11$, showing the aforementioned coexistence of particle- and hole-like ABS bands. 
 
The marked different behaviour of the subgap transport features confirms  that the operation regime of the device strongly depends on the geometric parameters. As already mentioned, the relevant  parameter controlling the position of the ABS is the dimensionless product  $k_F L_\text{M}$ \cite{Feijoo23_ABS_and_QPT_in_SE_SC_FMI_hybrids}. Although experimentally it would be difficult to have precise control over $L_\text{M}$, the Fermi momentum can be \textit{in situ} controlled via 
the back gate $V_\text{BG}$, providing a more accessible way to explore the quantum phase diagram of the device.

\subsection{Disorder effects}

In our previous calculations we have neglected the effects of disorder. This is a reasonable assumption in epitaxially grown interfaces (as is the case in our system of interest), where due to the controlled layer-by-layer deposition process, interfacial and structural disorder is minimized. In fact, the advent of this new generation of epitaxially-grown samples has dramatically improved the quality as compared to, e.g.,  SE-SC interfaces grown via evaporation techniques. \cite{Chang15_Hard_gap_in_SM_SC_NWs,Liu19_SM_FMI_SC_epitaxial_nanowires,Vaitiekenas21_ZBPs_in_FMI_SC_SM_hybrid_nanowires, Vaitiekenas22_Evidence_of_spin_polarized_ABS}. In particular, the ``soft gap" feature, characteristic of  SE nanowires with proximity-induced superconductivity grown using evaporation methods (see e.g. Refs. \cite{Mourik12_Signatures_of_MF, Das12_Evidence_of_MFs, Deng12_ZBP_in_Majorana_NW, RokhinsonNP, Churchill2013, Finck13_Majorana_wire} ), was virtually removed in this new generation of heterostructures.

However, disorder is a conceptually important effect and the question of how our previous results, obtained in the clean limit, could be affected by it remains an interesting and important question. In this Section we address the qualitative effects of disorder on the conductance and, in particular, we show that the presence of a moderate amount of static disorder does not significantly modify our main conclusions. The overall behavior displayed by the ABS in the disorder-free case, and the chemical potential values at which the quantum phase transition occurs remain practically unchanged.

We model the effects of disorder via local fluctuations in the chemical potential, substituting $\mu_\text{BG}\rightarrow\mu_\text{BG}+\delta\mu_j$ in Eq.(\ref{eq:H0}), as detailed in Ref.\cite{Fregoso13_Electrical_detection_of_TQPT}. We consider $\delta\mu_j$ as a $\delta$-correlated random variable, i.e., $\langle \delta \mu_i \delta \mu_j \rangle = v^2_0 \delta_{i,j}$, where the brackets correspond to the average over disorder realizations, and the parameter $v_0$ characterizes the overall ``disorder strength" in the system. In our calculations, we consider a specific disorder realization (i.e., a particular realization of the random disorder potential $\delta \mu_j$) and keep it fixed for the rest of our calculations. We next multiply $\delta \mu_j$ by  increasing  values of the parameter $v_0$ and we solve the corresponding eigenvalue equation Eq. (\ref{eq:eigenvalueBdG}). In Figure \ref{fig:disorder} we show the effect of increasing the disorder strength $v_0$ from the ``clean limit" regime $v_0/\Delta=0$ up to the moderate value $v_0/\Delta=0.6$. We note that the conductance peaks slightly deviate from the value associated with the clean limit case, without affecting the critical points where the ABS crossing at $\omega/\delta=0$ occurs. This plot indicates that the qualitative features of our results remain unchanged.

\section{Summary and conclusions}\label{sec:summary}

In this work we have studied the $T=0$ quantum transport properties of one-dimensional heterostructures realizing  semiconductor-superconductor-ferromagnetic insulator (SE-SC-FMI) triple interfaces. These devices have been recently fabricated using molecular beam-epitaxy growth techniques, and constitute an interesting example of a novel quantum device with tailored properties arising from the combination of materials. While most studies in these hybrid devices have focused on the potential generation of topological superconductor structures with Majorana fermions, in this work, we have proposed and studied a different application, i.e., the use of these hybrids to study spin-polarized transport in a highly tunable device.

We stress that the studied device differs from existing realizations in two fundamental aspects: a) the length $L_\text{M}$ of the FMI layer is shorter than the length of the SE-SC heterostructure, thereby introducing an inhomogeneous Zeeman interaction which has important effects on the induced ABS, and b) we consider SE nanowires with no (or very weak) Rashba spin-orbit interaction, a fact that allows to induce spin-polarized ABS and high-spin quantum ground states. Experimentally, the first requirement can be realized using appropriate masks (i.e., shadow-wall technique \cite{Carrad20_Shadow_Wall_Epitaxy_of_SE_SU_Hybrids, Heedt21_Shadow_Wall_Epitaxy_of_SE_SU_Hybrids}) in the MBE growth of the FMI layer. In addition, the second requirement could be accomplished by growing SE nanowires using centrosymmetric materials, such as silicon (Si). In practice, our results will remain unaffected whenever the residual spin-orbit coupling, proportional to a parameter  $\alpha$ (with units of energy) is such that $|\alpha|\ll \{\Delta, \gamma_R, \gamma_L\}$. Under such conditions, its effect will be unobservable and will not have measurable consequences for the nanowires considered here.

Finally, a crucial assumption consists in the possibility to modulate the total charge on the device via the presence of a uniform backgate $V_\text{BG}$. This point has been experimentally demonstrated in recent experiments with similar devices \cite{Vaitiekenas21_ZBPs_in_FMI_SC_SM_hybrid_nanowires, Vaitiekenas22_Evidence_of_spin_polarized_ABS}. Therefore we believe these three requirements are within experimental reach and could be well met in specific realizations.

Using reasonable values for the parameters in our model, our results indicate that controlling both $L_\text{M}$ and/or $V_\text{BG}$, the spin-polarized ABS can be tuned within the gap, and QPTs with abrupt changes in the fermion parity and in the total $S^z$ can be induced. As shown in Figs. \ref{fig:conductance vs energy 2} and \ref{fig:conductance for h=1.5 4L}, the zero-energy crossings of the subgap ABS as a function of $\mu_\text{BG}$ (via $V_\text{BG}$) would be an indication of such QPTs. We stress that using more involved experimental methods (i.e., Coulomb-blockade transport technique), tunable ABS crossing the Fermi energy have been already detected \cite{Vaitiekenas22_Evidence_of_spin_polarized_ABS}.

For completeness, it is worth discussing here the limit $k_\text{F}L_\text{M}\ll 1 $, in which the magnetic region becomes very small (much smaller than $\lambda_F$ the Fermi wavelength of the SE nanowire) in the presence of a very strong Zeeman field  $h_0$, such that  the product $L_\text{M}h_0 = J$ is kept constant. Under such conditions, the magnetic profile can be reasonably replaced by a delta function and the setup mimics the physics an atomic Yu-Shiba-Rusinov classical impurity  \cite{Yu65_YSR_states,Shiba68_YSR_states,Rusinov68_YSR_states}, in which the resulting ABS have been proposed as fundamental building blocks to engineer quantum devices with topologically non-trivial ground states \cite{Nadj-Perdge13_Majorana_fermions_in_Shiba_chains,
Klinovaja13_TSC_and_Majorana_Fermions_in_RKKY_Systems, Li14_TSC_induced_by_FM_metal_chains}.  In this case, the ABS with odd spatial symmetry with respect to the center of the magnetic profile decouple, as they have vanishing weight at the magnetic region,  and only ABS with even symmetry survive in the subgap region. The theoretical prediction in that case is that a parity- and spin-changing QPT would occur when $J$ satisfies the condition $\rho_0 J=1$, with $\rho_0$ the density of normal conduction states at the Fermi level \cite{Yu65_YSR_states,Shiba68_YSR_states,Rusinov68_YSR_states}. From that perspective, our results are a generalization of the impurity case to situations where the magnetic region is extended, and therefore more QPTs are expected to occur (see also  Ref. \cite{Feijoo23_ABS_and_QPT_in_SE_SC_FMI_hybrids}).

We now briefly discuss the case of finite temperature. Although the main quantities discussed in this work, in particular the quantum phase diagram and the QPTs, are well-defined only at $T=0$, the presence of the induced SC gap $\Delta$ makes all these quantities well-behaved in the case $k_B T\ll \Delta $ and smoothly connected to the $T=0$ case. This has important implications of real experiments, which are always done at a finite temperature. From  Eqs. (\ref{eq:G_LL_final3}) and (\ref{eq:G_LR_final3}), we note that the derivative of the Fermi function, which at $T=0$ reduces to a Dirac delta function centered at $\omega=eV_L$, becomes broadened at finite temperature, with a characteristic width of order $k_B T$, and therefore  the integral effectively turns into a convolution with a thermal ``smearing'' function $d n_F(\omega-V_L) / dV_L$.  To quantify thermal effects, in the Appendix \ref{sec:finiteT} we have explicitly calculated the local conductance $\mathsf{G}_{LL}$  for a system with $N_\text{M}=38$, for two different parameters $\mu_\text{BG}/\Delta=-2$ and $\mu_\text{BG}/\Delta=-8$, and for different temperatures. Thermal smearing effects on the conductance peaks are clearly visible in Fig. (\ref{fig:thermal_smearing}), which makes the identification of the zero-crossing point associated with the phase transition more challenging. However, we stress that the position of the zero-crossing point does not change in relation to the parameter $\mu_\text{BG}/\Delta$, and therefore our main conclusions (e.g., those in Fig. \ref{fig:conductance for h=1.5 4L}) are not qualitatively affected.

Finally, we stress that in our calculations we have neglected other experimentally relevant effects:  multiple conduction channels, electron-electron interaction, etc. While our results will certainly be modified if these effects are taken into account, we expect they will not affect our conclusions qualitatively, in particular the existence of parity- and spin-changing QPTs in this type of devices, and the possibility of their experimental detection. A notable exception is the presence of residual spin-orbit coupling producing a mixing of  ABS branches with different spin projections, which would lift the degeneracy whenever two such branches become degenerate. In particular, spin-orbit coupling would generate avoided crossings instead of zero-energy crossings, which would modify the properties of the QPTs discussed in this work. However, we believe that this particular effect would be minimized using centrosymmetric SE materials, such as silicon, instead of InAs. In conclusion, we believe the device proposed in this work is within experimental reach using present-day nanofabrication techniques and could 
have important potential uses in low-dissipation spin-polarized transport.

\acknowledgments{This work is partially supported by CONICET, Universidad Nacional de La Plata (UNLP) and Universidad Nacional de Cuyo (UNCuyo), Argentina.}

\section*{Declarations}

\paragraph*{\textbf{Author Contribution Statement:}} Anibal Iucci and Alejandro M. Lobos, conceived the theoretical model and developed the approximation strategies. Javier S. Feijóo carried out all numerical calculations and generated the figures. All authors participated in the discussions, contributed equally to the final manuscript, and approved its submission.
\\
\paragraph*{\textbf{Data Availability Statement:}} 
This study does not include any external datasets. All plots and figures can be fully reproduced using the analytical expressions provided in the main text and appendices.

\appendix

\section{Coupling to the measuring circuits}\label{app:coupling_to_the_leads}

Following standard procedures, the Green's function of the coupled system central region + measuring contacts  can be obtained from the Dyson's equation \cite{Haug_Jauho_book}:
\begin{align} \label{eq:Dyson_lead}
\hat{\mathbf{g}}^{r}_{i,j;\sigma}(\omega)&=\hat{\mathbf{g}}^{(0),r}_{i,j;\sigma}(\omega)+\hat{\mathbf{g}}_{i,1;\sigma}^{(0),r}(\omega)\hat{\boldsymbol{\Sigma}}_L(\omega)\hat{\mathbf{g}}_{1,j;\sigma}^{r}(\omega)\nonumber \\
&+\hat{\mathbf{g}}_{i,N;\sigma}^{(0),r}(\omega)\hat{\boldsymbol{\Sigma}}_R(\omega)\hat{\mathbf{g}}_{N,j;\sigma}^{r}(\omega)
\end{align}
where
\begin{equation}
\hat{\boldsymbol{\Sigma}}_{\alpha;\sigma}(\omega)\equiv\hat{\mathbf{t}}_{\alpha}\hat{\mathbf{g}}_{\alpha,\alpha;\sigma}^{(0),r}(\omega)\hat{\mathbf{t}}^*_{\alpha},
\end{equation}
with $\alpha=\{L,R\}$, and where
\begin{equation}
\hat{\mathbf{t}}_\alpha =\begin{pmatrix}
   t_\alpha  & 0 \\
    0 & -t_\alpha
\end{pmatrix}
\end{equation}
is a $2 \times 2$ matrix containg the hopping parameters with the leads, and 
\begin{equation}
\hat{\mathbf{g}}^{(0),r}_{\alpha,\alpha;\sigma}(\omega) =\begin{pmatrix}
  g^{(0),r}_{\alpha,\alpha;\sigma}(\omega)  & 0 \\
    0 & \bar{g}^{(0),r}_{\alpha,\alpha;\sigma}(\omega)
\end{pmatrix},
\end{equation}
is the Green's function of the leads before coupling to the sites  $j=\{1,N\}$  of the SE-SC-FMI heteroestructure.
The Dyson's equation Eq. (\ref{eq:Dyson_lead}) can be more compactly expressed in terms of the $2N \times  2N$ matrix equation
\begin{align}\label{eq:Dyson_compact}
\hat{\mathbf{g}}_{\sigma}^{r}(\omega)&=\hat{\mathbf{g}}_{\sigma}^{(0),r}(\omega)+\hat{\mathbf{g}}_{\sigma}^{(0),r}(\omega)\hat{\boldsymbol{\Sigma}}_{\sigma}(\omega)\hat{\mathbf{g}}_{\sigma}^{r}(\omega),
\end{align}
where the resolvent matrix $\hat{\mathbf{g}}^{(0),r}(\omega)$ can be conveniently expressed in terms of the eigenmodes in Eq. (\ref{eq:eigenvectors}) as
\begin{align}
\hat{\mathbf{g}}_\sigma^{(0),r}(\omega)&=\left[\omega +i\eta - \mathcal{H}^\text{w}_{0,\sigma}\right]^{-1},\nonumber\\
&=\sum_\nu \frac{\Phi_{\nu,\sigma}\cdot\left(\Phi_{\nu,\sigma}\right)^\dagger}{\omega -\epsilon_{\nu,\sigma}+i \eta},
\end{align}
and the self-energy matrix $\hat{\boldsymbol{\Sigma}}(\omega)$ is defined as
\begin{align}
\left[\hat{\boldsymbol{\Sigma}}_\sigma(\omega)\right]_{i,j}&=t_L^2\left[g^{(0),r}_{L,L}(\omega)\delta_{i,1}\delta_{j,1}\right.\nonumber\\
&\qquad+\left.\bar{g}^{(0),r}_{L,L}(\omega)\delta_{i,N+1}\delta_{j,N+1}\right]\nonumber\\
&+t_R^2\left[g^{(0),r}_{R,R}(\omega)\delta_{i,N}\delta_{j,N}\right.\nonumber\\
&\qquad+\left.\bar{g}^{(0),r}_{R,R}(\omega)\delta_{i,2N}\delta_{j,2N}\right].
\end{align}
Note that Eq. (\ref{eq:Dyson_compact}) is a linear matrix equation for  $\hat{\mathbf{g}}^{r}_{\sigma}(\omega)$ which in principle could be solved performing a matrix inversion
\begin{align}\label{eq:gr_inversion}
\hat{\mathbf{g}}^{r}_{\sigma}(\omega)&=\left[\omega +i\eta  - \mathcal{H}^\text{w}_{0,\sigma}-\hat{\boldsymbol{\Sigma}}_{\sigma}(\omega)\right]^{-1}.
\end{align}
However, this is a cumbersome procedure from the numerical point of view, since it involves a matrix inversion for each value of $\omega$. To avoid this, we assume  the  wide-band limit in the normal leads, in which case $g^{(0),r}_{\alpha,\alpha;\sigma}(\omega)\rightarrow-i\pi\rho_{\alpha}^{(0)}$ where $\rho_{\alpha}^{(0)}$ is the density of states at the Fermi energy in lead $\alpha$ (note that we have omitted the spin indices here assuming SU(2) symmetry in the leads). Therefore  the matrix $\hat{\boldsymbol{\Sigma}}_{\sigma}(\omega)=\hat{\boldsymbol{\Sigma}}(\omega)$ becomes static and can be expressed as
$\hat{\boldsymbol{\Sigma}}(\omega)\rightarrow -i\hat{\boldsymbol{\Gamma}}$, with
\begin{align}
\left[\hat{\boldsymbol{\Gamma}}\right]_{i,j}&= \gamma_L\left(\delta_{i,1}\delta_{j,1}+\delta_{i,N+1}\delta_{j,N+1}\right)\nonumber\\
&+\gamma_R\left(\delta_{i,N}\delta_{j,N}+\delta_{i,2N}\delta_{j,2N}\right),
\end{align}
where we have defined the effective hybridization parameter $\gamma_\alpha=\pi t^2_\alpha \rho^{(0)}_\alpha$. 

Then, the  Green's function in Eq. (\ref{eq:gr_inversion}) can be obtained solving the new eigenvalue problem with non-hermitian matrices \cite{datta}
\begin{align}
\left[\mathcal{H}^\text{w}_{0,\sigma}-i\hat{\Gamma}\right]\boldsymbol{\chi}
_{\nu,\sigma}&=\tilde{\epsilon}_{\nu,\sigma}\boldsymbol{\chi}_{\nu,\sigma}\\
\left[\mathcal{H}^\text{w}_{0,\sigma}+i\hat{\Gamma}\right]\boldsymbol{\chi}^*_{\nu,\sigma}&=\tilde{\epsilon}^*_{\nu,\sigma}\boldsymbol{\chi}^*_{\nu,\sigma},
\label{eq:complex_eigensystem}
\end{align} 
where the eigenvalues $\tilde{\epsilon}_{\nu,\sigma}$ are complex, and the $2N-$component eigenvectors $\boldsymbol{\chi}_{\nu,\sigma}= \left(\tilde{\mathbf{u}}_{\nu,\sigma},\tilde{\mathbf{v}}_{\nu,\sigma}\right)^T$ obey the orthogonality condition  $ 
\boldsymbol{\chi}^T_{\mu,\sigma}\cdot\boldsymbol{\chi}_{\nu,\sigma}=\delta_{\mu,\nu}$. These eigenvectors can be used to diagonalize the non-hermitian matrix $\mathcal{H}^\text{w}_{0,\sigma}+i\hat{\Gamma}$, and therefore the Green's function of the SE-SC-FMI structure dressed with the leads can be computed  as
\begin{equation} \label{eq:Gtotaleigen}
\hat{\mathbf{g}}_{\sigma}^{r}(\omega)=\sum_\nu \frac{\boldsymbol{\chi}_{\nu,\sigma}\cdot\left(\boldsymbol{\chi}_{\nu,\sigma}\right)^T}{\omega -\tilde{\epsilon}_{\nu,\sigma}+i \eta}.
\end{equation}
Whenever necessary, the spatial projection onto sites $\{i,j\} \in \{1,\dots, N\}$ is simply
\begin{align}    \hat{\mathbf{g}}^{r}_{i,j;\sigma}(\omega)&=\sum_\nu \frac{1}{\omega-\tilde{\epsilon}_{\nu,\sigma}+i\eta} \left(
\begin{array}{cc}
\tilde{u}^\sigma_{i,\nu}\left(\tilde{u}^\sigma_{j,\nu}\right) &\tilde{u}^\sigma_{i,\nu}\left(\tilde{v}^\sigma_{j,\nu}\right)\\
\tilde{v}^\sigma_{i,\nu}\left(\tilde{u}^\sigma_{j,\nu}\right)&\tilde{v}^\sigma_{i,\nu}\left(\tilde{v}^\sigma_{j,\nu}\right)
\end{array}
\right).\label{eq:Gtotaleigen_spatial_proj}
\end{align}
\begin{widetext}
\section{Embedding procedure}\label{app:embedding}

A useful trick to eliminate finite-size effects in our calculations consists in coupling the Green's function in Eq. (\ref{eq:Gtotaleigen}), already dressed with the contacts, to  the semi-infinite chains defined in Eqs. (\ref{eq:HwL}) and (\ref{eq:HwR}). Of course, this is a method valid only in non-interacting models, as is the case here. We therefore introduce the following Dyson's equation
\begin{align}
\hat{\mathbf{G}}^{r}_\sigma(\omega)&=\hat{\mathbf{G}}^{(0),r}_\sigma(\omega)+\hat{\mathbf{G}}^{(0),r}_\sigma(\omega)\mathcal{H}^\text{w}_\text{mix}\hat{\mathbf{G}}^{r}_\sigma(\omega),\label{eq:Dyson_embedding}
\end{align}
where we have defined the (formally infinite) decoupled Green's function matrix as $\hat{\mathbf{G}}^{(0),r}_\sigma(\omega)\equiv\hat{\mathbf{g}}^{(0),r}_{\text{w},L;\sigma}(\omega)\otimes \hat{\mathbf{g}}^{r}_\sigma(\omega)\otimes \hat{\mathbf{g}}^{(0),r}_{\text{w},R;\sigma}(\omega)$, which encompasses the three regions defined by Eqs. (\ref{eq:H0}), (\ref{eq:HwL}) and (\ref{eq:HwR}). In this expression, the Green's function of the central region $\hat{\mathbf{g}}^{r}_\sigma(\omega)$ corresponds to Eq. (\ref{eq:Gtotaleigen}), while the other Green's functions correspond to those of the semi-infinite SC-SE chains. The matrix $\mathcal{H}^\text{w}_\text{mix}$ is the Nambu matrix corresponding to the Hamiltonian Eq. (\ref{eq:Hmix_wire}). Finally, $\hat{\mathbf{G}}^{r}_\sigma(\omega)$ is the fully dressed Green's function matrix of the problem, which we need to compute in order to obtain the transport properties of the device (see next Section).
Since the calculation of transport properties involves the computation of   $\hat{\mathbf{G}}^{r}_\sigma(\omega)$ at sites belonging to the range $\{i,j \}\in \{1, 2,\dots, N\}$, we project the above Dyson's equation in real space:
\begin{align}
\hat{\mathbf{G}}^{r}_{i,j;\sigma}(\omega)&=\hat{\mathbf{g}}^{r}_{i,j;\sigma}+\hat{\mathbf{g}}^{r}_{i,1;\sigma}\mathcal{H}^\text{w}_{\text{mix};\sigma}\hat{\mathbf{g}}^{(0),r}_{0,0;\sigma}\mathcal{H}^\text{w}_{\text{mix};\sigma}\hat{\mathbf{G}}^{r}_{1,j;\sigma}
+\hat{\mathbf{g}}^{r}_{i,N;\sigma}\mathcal{H}^\text{w}_{\text{mix};\sigma}\hat{\mathbf{g}}^{(0),r}_{N+1,N+1;\sigma}\mathcal{H}^\text{w}_{\text{mix};\sigma}\hat{\mathbf{G}}^{r}_{N,j;\sigma} \label{eq:Dyson_G_final}\\[1 em]
&\equiv\begin{pmatrix}
    G^r_{i,j;\sigma}(\omega)&F^r_{i,j;\sigma}(\omega)  \\     \bar{F}^r_{i,j;\sigma}(\omega)&\bar{G}^r_{i,j;\sigma}(\omega) 
    \end{pmatrix}
 \hspace{2cm}\text{for\  }\{i,j\}=\{1,2, \dots, N\},\label{eq:G_fully_dressed}
\end{align}
where in the first line we have omitted the frequency dependence and where we have defined the Nambu matrix corresponding to Eq. (\ref{eq:Hmix_leads})  as
 \begin{equation}
\mathcal{H}^\text{w}_{\text{mix};\sigma}=\begin{pmatrix}
   t  & 0 \\
     0 & -t
 \end{pmatrix}.\label{eq:Hwmix}
 \end{equation}
In addition, the matrices $\hat{\mathbf{g}}^{(0),r}_{0,0;\sigma}(\omega)$ and $\hat{\mathbf{g}}^{(0),r}_{N+1,N+1;\sigma}(\omega)$ are the Nambu Green's functions corresponding to the first sites of the semi-infinite chains, which can be obtained via the recursive method, and whose concrete expression is given  Appendix \ref{Ap:semi-infinite}. We have checked that embedding the central region in this way effectively eliminates finite-size effects in the different computed quantities (e.g., position of the intragap ABS). Note that Eq. (\ref{eq:Dyson_G_final}) is a system of coupled equations for the four  2$\times$2 matrices $\hat{\mathbf{G}}^{r}_{1,1;\sigma}(\omega)$, $\hat{\mathbf{G}}^{r}_{1,N;\sigma}(\omega)$, $\hat{\mathbf{G}}^{r}_{N,1;\sigma}(\omega)$, and $\hat{\mathbf{G}}^{r}_{N,N;\sigma}(\omega)$, which we have solved quasi-analytically, and we give the details in the Appendix \ref{Ap:embedding}.

\subsection{Local Nambu Green's function of a semi-infinite chain} \label{Ap:semi-infinite}
In this section, we derive the expressions for $\hat{\mathbf{g}}^{(0),r}_{0,0}(\omega)$ and $\hat{\mathbf{g}}^{(0),r}_{N+1,N+1}(\omega)$ that we need in order to couple the semi-infinite chains to the central region via Eq. (\ref{eq:Dyson_G_final}). For simplicity, we describe here a generic semi-infinite chain whose site labels range from $j=1$ to $j=\infty$ with the following Hamiltonian
\begin{align}
    H^\text{w}_\text{SI}&=\sum_{j=1,\sigma}^{\infty} \left[-\mu_\text{BG}  c_{j,\sigma}^{\dagger}c_{j,\sigma} -t\left(c_{j,\sigma}^{\dagger}c_{j+1,\sigma}+\textrm{H.c.}\right)\right]\nonumber\\
&+\Delta\sum_{j=1}^{\infty}\left(c_{j,\uparrow}^{\dagger}c_{j,\downarrow}^{\dagger}+\textrm{H.c.} \right)\label{eq:SI_appendix},  
\end{align}
 which we can identify with the semi-infinite Hamiltonians in Eqs. (\ref{eq:HwL}) and (\ref{eq:HwR}). Since the corresponding Green's function matrix is formally infinite, we define the spatially-projected retarded Green's function matrix on sites $j,l$  in Nambu space as 
\begin{align}
\hat{\mathbf{g}}^{\text{SI}}_{j,l}(\omega)&=\left[\omega +i\eta -\hat{\mathcal{H}}^\text{w}_\text{SI}\right]^{-1}_{j,l},
\end{align}
where $\hat{\mathcal{H}}^\text{w}_\text{SI}$ is the (infinite) BdG Hamiltonian matrix corresponding to Eq. (\ref{eq:SI_appendix}).

We now introduce a new site $j=0$, identical to all other sites in Eq. (\ref{eq:SI_appendix}), to the left of the chain. It is evident that the system remains the same (i.e., it is a semi-infinite chain). The BdG Hamiltonian of this new site in the basis $\left(c_{0,\uparrow},c_{0,\downarrow}^{\dagger}\right)^T$ is 
\begin{equation}
    \hat{\mathcal{H}}_{0,0}=\begin{pmatrix}
    -\mu_\text{BG} & \Delta\\
    \Delta & \mu_\text{BG}
    \end{pmatrix}, \label{eq:h0_site0}
\end{equation}      
and it couples to the site $j=1$ through the term (see Eq. (\ref{eq:Hwmix}))
\begin{align}
\left(\hat{\mathcal{H}}^\text{w}_{\text{mix}}\right)_{j,l}&=t \hat{\tau_{z}}\left(\delta_{j,1}\delta_{l,0}+\delta_{j,0}\delta_{l,1}\right),\label{eq:H_mix_BdG}
\end{align}
where $\hat{\tau}_z$ is the $2\times 2$ Pauli matrix along $z$. Note that we have dropped the spin index in the above expressions as these portions of the nanowire are SU(2) symmetric. Since we are interested in the local Green's function $\hat{\mathbf{g}}^{\text{SI}}_{0,0}(\omega)$, we can use  Dyson's equation:
\begin{align}
  \hat{\mathbf{g}}^{\text{SI}}_{0,0}(\omega) &= \hat{\mathbf{g}}^{\text{SI},(0)}_{0,0}(\omega) + \hat{\mathbf{g}}^{\text{SI},(0)}_{0,0}(\omega)\hat{\mathcal{H}}^\text{w}_{\text{mix}}  \hat{\mathbf{g}}^{\text{SI}}_{1,0}(\omega) \label{eq:g00}\\
 \hat{\mathbf{g}}^{\text{SI}}_{1,0}(\omega) &=  \hat{\mathbf{g}}^{\text{SI},(0)}_{1,1}(\omega)\hat{\mathcal{H}}^\text{w}_{\text{mix}} \hat{\mathbf{g}}^{\text{SI}}_{0,0}(\omega)\label{eq:g10},    
\end{align}
where $ \hat{\mathbf{g}}^{\text{SI},(0)}_{0,0}(\omega)$ is the Nambu Green's corresponding to the isolated site $j=0$
\begin{align}
   \hat{\mathbf{g}}^{\text{SI},(0)}_{0,0}(\omega) &= \left[\omega+i\eta- \mathcal{H}_{0,0}\right]^{-1}\nonumber\\
    &= \frac{1}{D\left(\omega\right)}\begin{pmatrix}
    \omega+i\eta-\mu_\text{BG} & \Delta\\
    \Delta & \omega+i\eta+\mu_\text{BG}
    \end{pmatrix},
\end{align}
where $D\left(\omega\right)=(\omega+i\eta)^{2}-\mu_\text{BG}^{2}-\Delta^{2}$.    
Replacing Eq. (\ref{eq:g10}) into Eq. (\ref{eq:g00}), and noticing that $  \hat{\mathbf{g}}^{\text{SI}}_{0,0}(\omega)=   \hat{\mathbf{g}}^{\text{SI,(0)}}_{1,1}(\omega)$, we find a closed  equation for the matrix $\hat{\mathbf{g}}^{\text{SI}}_{0,0}(\omega)$:  
\begin{equation}
  \hat{\mathbf{g}}^{\text{SI}}_{0,0}(\omega) = \hat{\mathbf{g}}^{\text{SI},(0)}_{0,0}(\omega) +\hat{\mathbf{g}}^{\text{SI},(0)}_{0,0}(\omega)
\hat{\mathcal{H}}^\text{w}_{\text{mix}}  \hat{\mathbf{g}}^{\text{SI}}_{0,0}(\omega)\hat{\mathcal{H}}^\text{w}_{\text{mix}} \hat{\mathbf{g}}^{\text{SI}}_{0,0}(\omega)\label{eq:recursive_green}.  
\end{equation}
The Eq. (\ref{eq:recursive_green}) defines a coupled system of quadratic equations for the 4 Nambu elements of $\hat{\mathbf{g}}^{\text{SI}}_{0,0}(\omega)$ that can be solved analytically: 

    \begin{align}
         g^{\text{SI}}_{0,0}&=\frac{1}{D\left(\omega\right)}\left[\left(\omega+i\eta-\mu_\text{BG}\right)+\left|t\right|^{2}\left[\left(\left(g^{\text{SI}}_{0,0}\right)^2-f^{\text{SI}}_{0,0}\bar{f}^{\text{SI}}_{0,0}\right)\left(\omega+i\eta-\mu_\text{BG}\right)+\Delta\left(\bar{g}^{\text{SI}}_{0,0}\bar{f}^{\text{SI}}_{0,0}-\bar{f}^{\text{SI}}_{0,0}g^{\text{SI}}_{0,0}\right)\right]\right],\\
         f^{\text{SI}}_{0,0}&=\frac{1}{D\left(\omega\right)}\left[\Delta+\left|t\right|^{2}\left[\left(g^{\text{SI}}_{0,0}f^{\text{SI}}_{0,0}-f^{\text{SI}}_{0,0}\bar{g}^{\text{SI}}_{0,0}\right)\left(\omega+i\eta-\mu_\text{BG}\right)+\Delta\left(\left(\bar{g}^{\text{SI}}_{0,0}\right)^2-\bar{f}^{\text{SI}}_{0,0}f^{\text{SI}}_{0,0}\right)\right]\right],\\
         \bar{f}^{\text{SI}}_{0,0}&=\frac{1}{D\left(\omega\right)}\left[\Delta+\left|t\right|^{2}\left[\left(\bar{g}^{\text{SI}}_{0,0}\bar{f}^{\text{SI}}_{0,0}-\bar{f}^{\text{SI}}_{0,0}g^{\text{SI}}_{0,0}\right)\left(\omega+i\eta+\mu_\text{BG}\right)+\Delta\left(\left(g^{\text{SI}}_{0,0}\right)^2-f^{\text{SI}}_{0,0}\bar{f}^{\text{SI}}_{0,0}\right)\right]\right],\\
         \bar{g}^{\text{SI}}_{0,0}&=\frac{1}{D\left(\omega\right)}\left[\left(\omega+i\eta+\mu_\text{BG}\right)-\left|t\right|^{2}\left[\left(\left(\bar{g}^{\text{SI}}_{0,0}\right)^2-\bar{f}^{\text{SI}}_{0,0}f^{\text{SI}}_{0,0}\right)\left(\omega+i\eta+\mu_\text{BG}\right)+\Delta\left(g^{\text{SI}}_{0,0}f^{\text{SI}}_{0,0}-f^{\text{SI}}_{0,0}\bar{g}^{\text{SI}}_{0,0}\right)\right]\right],\label{eq:g0_semi_infinite}
 \end{align}
where for compactness the frequency argument has been avoided.  These expressions can be simplified to
    \begin{align}
         g^{\text{SI}}_{0,0}(\omega)&=\frac{1}{4t^{2}}\left[2\left(\omega+i\eta+\mu_\text{BG}\right)\pm\sqrt{2}\frac{\left(2\left(\omega+i\eta\right)\mu_\text{BG}+K_{\pm}(\omega)\right)}{\sqrt{K_{\pm}(\omega)}}\right],\\
         f^{\text{SI}}_{0,0}(\omega)&=\frac{\Delta}{2t^{2}}\left(1\pm\frac{\mu_\text{BG}\sqrt{2}}{\sqrt{K_{\pm}(\omega)}}\right),\\
         \bar{f}^\text{SI}_{0,0}(\omega)&=\frac{\Delta}{2t^{2}}\left(1\pm\frac{\mu_\text{BG}\sqrt{2}}{\sqrt{K_{\pm}(\omega)}}\right),\\
         \bar{g}^{\text{SI}}_{0,0}(\omega)&=\frac{1}{4t^{2}}\left[2\left(\omega+i\eta-\mu_\text{BG}\right)\pm\sqrt{2}\frac{\left(2\left(\omega+i\eta\right)\mu_\text{BG}-K_{\pm}(\omega)\right)}{\sqrt{K_{\pm}(\omega)}}\right],\label{eq:g0_semi_infinite2}
    \end{align}

where we have defined $K_{\pm}(\omega)=T(\omega) \pm \sqrt{S(\omega)}$, and 
\begin{align}
    T(\omega)&=-4t^{2}+\left(\omega+i\eta\right)^{2}-\Delta^{2}+\mu_\text{BG}^{2}\\
    S(\omega)&=16t^{4}-8t^{2}\left(\left(\omega+i\eta\right)^{2}-\Delta^{2}+\mu_\text{BG}^{2}\right)+\left(\left(\omega+i\eta\right)^{2}-\Delta^{2}-\mu_\text{BG}^{2}\right)^{2}.
\end{align}
In the above expressions, the sign $\pm$ is selected imposing continuity conditions for $\omega$, and requiring that $\text{Im}\left[\hat{g}^{\text{SI}}_{0,0}(\omega)\right]<0$. Finally, the Green's functions needed in Eq. (\ref{eq:Dyson_G_final}) are
\begin{align}
\hat{\mathbf{g}}^{(0),r}_{0,0}(\omega)&=\hat{\mathbf{g}}^{(0),r}_{N+1,N+1}(\omega)=\hat{\mathbf{g}}^{\text{SI}}_{0,0}(\omega).\label{eq:semi_infinite_final}
\end{align}

\subsection{Embedding the SE-SC-FMI finite system via the Dyson's equation}\label{Ap:embedding}
In order to eliminate finite-size effects we couple the central region in Fig. \ref{fig:model scheme} (containing the SE-SC-FMI hybrid structure already connected to the contacts Eq. (\ref{eq:Gtotaleigen})) to semi-infinite SE-SC chains given by Eqs. (\ref{eq:HwL}) and (\ref{eq:HwR}). The term that couples the finite system to the semi-infinite chains is Eq. (\ref{eq:Hmix_wire}). Therefore the Dyson's equation that dresses the central part is

\begin{align}
\hat{\mathbf{G}}^{r}_\sigma(\omega)&=\hat{\mathbf{G}}^{(0),r}_\sigma(\omega)+\hat{\mathbf{G}}^{(0),r}_\sigma(\omega)\mathcal{H}^\text{w}_\text{mix}\hat{\mathbf{G}}^{r}_\sigma(\omega) \label{eq:Gembed}
\end{align}

This is a formally infinite matrix equation, which can be projected onto the sites corresponding to the central region, and  solved quasi-analytically for the site-projections of Eq. (\ref{eq:Gembed}) that are needed for the transport properties [i.e., Eqs. (\ref{eq:conductance_GLL})-(\ref{eq:conductance_GRL})]

\begin{align}
\hat{\mathbf{G}}^{r}_{i,j;\sigma}(\omega)&=\hat{\mathbf{g}}^{r}_{i,j;\sigma}+\hat{\mathbf{g}}^{r}_{i,1;\sigma}\mathcal{H}^\text{w}_{\text{mix}}\hat{\mathbf{g}}^{(0),r}_{0,0;\sigma}\mathcal{H}^\text{w}_{\text{mix}}\hat{\mathbf{G}}^{r}_{1,j;\sigma}
+\hat{\mathbf{g}}^{r}_{i,N;\sigma}\mathcal{H}^\text{w}_{\text{mix}}\hat{\mathbf{g}}^{(0),r}_{N+1,N+1;\sigma}\mathcal{H}^\text{w}_{\text{mix}}\hat{\mathbf{G}}^{r}_{N,j;\sigma} \label{eq:Dyson_G_final2}\\
&\equiv\begin{pmatrix}
    G^r_{i,j;\sigma}(\omega)&F^r_{i,j;\sigma}(\omega)  \\     \bar{F}^r_{i,j;\sigma}(\omega)&\bar{G}^r_{i,j;\sigma}(\omega) 
\end{pmatrix}, \hspace{2cm}\text{for\  }\{i,j\}=\{1,N\},\label{eq:G_fully_dressed2}
\end{align}

where in the first line we have omitted the frequency argument for compactness, and where we have used the result  $\hat{\mathbf{g}}^{(0),r}_{0,0;\sigma}(\omega) = \hat{\mathbf{g}}^{(0),r}_{N+1,N+1;\sigma}(\omega)=\hat{\mathbf{g}}^{\text{SI}}_{0,0}(\omega)$, obtained in Eq. (\ref{eq:semi_infinite_final}). We only need 4 projections of the fully dressed Green's functions to obtain the local and non-local conductance [see Sec.(\ref{sec:transport_properties})]

\begin{align}
\hat{\mathbf{G}}^{r}_{1,1;\sigma}(\omega)&=\hat{\mathbf{g}}^{r}_{1,1;\sigma}+\left|t\right|^{2}\left[\hat{\mathbf{g}}^{r}_{1,1;\sigma}\hat{\Omega}_{1,1}\hat{\mathbf{G}}^{r}_{1,1;\sigma}+\hat{\mathbf{g}}^{r}_{1,N;\sigma}\hat{\Omega}_{N,N}\hat{\mathbf{G}}^{r}_{N,1;\sigma}\right], \label{Eq:G11} \\ 
\hat{\mathbf{G}}^{r}_{1,N;\sigma}(\omega)&=\hat{\mathbf{g}}^{r}_{1,N;\sigma}+\left|t\right|^{2}\left[\hat{\mathbf{g}}^{r}_{1,1;\sigma}\hat{\Omega}_{1,1}\hat{\mathbf{G}}^{r}_{1,N;\sigma}+\hat{\mathbf{g}}^{r}_{1,N;\sigma}\hat{\Omega}_{N,N}\hat{\mathbf{G}}^{r}_{N,N;\sigma}\right], \label{Eq:G1N} \\
\hat{\mathbf{G}}^{r}_{N,1;\sigma}(\omega)&=\hat{\mathbf{g}}^{r}_{N,1;\sigma}+\left|t\right|^{2}\left[\hat{\mathbf{g}}^{r}_{N,1;\sigma}\hat{\Omega}_{1,1}\hat{\mathbf{G}}^{r}_{1,1;\sigma}+\hat{\mathbf{g}}^{r}_{N,N;\sigma}\hat{\Omega}_{N,N}\hat{\mathbf{G}}^{r}_{N,1;\sigma}\right], \label{Eq:GN1} \\ 
\hat{\mathbf{G}}^{r}_{N,N;\sigma}(\omega)&=\hat{\mathbf{g}}^{r}_{N,N;\sigma}+\left|t\right|^{2}\left[\hat{\mathbf{g}}^{r}_{N,1;\sigma}\hat{\Omega}_{1,1}\hat{\mathbf{G}}^{r}_{1,N;\sigma}+\hat{\mathbf{g}}^{r}_{N,N;\sigma}\hat{\Omega}_{N,N}\hat{\mathbf{G}}^{r}_{N,N;\sigma}\right], \label{Eq:GNN} 
\end{align}
where we have defined the quantities
\begin{align}
    \hat{\Omega}_{j}&=\hat{\tau}_{z}\hat{\mathbf{g}}^\text{SI}_{j,j}\hat{\tau}_{z},\quad\quad \text{for }j=\{1,N\}.
\end{align}
These dressed Green's functions form a linear set of equations that can be solved analytically. Solving first for $\hat{\mathbf{G}}^{r}_{N,1;\sigma}(\omega)$ and $\hat{\mathbf{G}}^{r}_{N,N;\sigma}(\omega)$ from Eqs. (\ref{Eq:GN1}) and Eq. (\ref{Eq:GNN}) respectively 

\begin{align}
\hat{\mathbf{G}}^{r}_{N,1;\sigma}(\omega)&=\left(\mathbb{1}-\left|t\right|^{2}\hat{\mathbf{g}}^{r}_{N,N;\sigma}\hat{\Omega}_{N,N}\right)^{-1}\left(\hat{\mathbf{g}}^{r}_{N,1;\sigma}+\left|t\right|^{2}\hat{\mathbf{g}}^{r}_{N,1;\sigma}\hat{\Omega}_{1,1}\hat{\mathbf{G}}^{r}_{1,1;\sigma}\right), \\
\hat{\mathbf{G}}^{r}_{N,N;\sigma}(\omega)&=\left(\mathbb{1}-\left|t\right|^{2}\hat{\mathbf{g}}^{r}_{N,N;\sigma}\hat{\Omega}_{N,N}\right)^{-1}\left(\hat{\mathbf{g}}^{r}_{N,N;\sigma}+\left|t\right|^{2}\hat{\mathbf{g}}^{r}_{N,1;\sigma}\hat{\Omega}_{1,1}\hat{\mathbf{G}}^{r}_{1,N;\sigma}\right),
\end{align}
and replacing into Eqs. (\ref{Eq:G11}) and (\ref{Eq:G1N}), we reduce the unknowns to two independent Green's functions 
 
\begin{align}
    \hat{\mathbf{G}}^{r}_{1,1;\sigma}(\omega)&=\hat{\mathbf{g}}^{r}_{1,1;\sigma}+\left|t\right|^{2}\left[\hat{\mathbf{g}}^{r}_{1,1;\sigma}\hat{\Omega}_{1,1}\hat{\mathbf{G}}^{r}_{1,1;\sigma}+\hat{\mathbf{g}}^{r}_{1,N;\sigma}\hat{\Omega}_{N,N}\left(\mathbb{1}-\left|t\right|^{2}\hat{\mathbf{g}}^{r}_{N,N;\sigma}\hat{\Omega}_{N,N}\right)^{-1}\left(\hat{\mathbf{g}}^{r}_{N,1;\sigma}+\left|t\right|^{2}\hat{\mathbf{g}}^{r}_{N,1;\sigma}\hat{\Omega}_{1,1}\hat{\mathbf{G}}^{r}_{1,1;\sigma}\right)\right], \label{Eq:G11_s} \\ \hat{\mathbf{G}}^{r}_{1,N;\sigma}(\omega)&=\hat{\mathbf{g}}^{r}_{1,N;\sigma}+\left|t\right|^{2}\left[\hat{\mathbf{g}}^{r}_{1,1;\sigma}\hat{\Omega}_{1,1}\hat{\mathbf{G}}^{r}_{1,N;\sigma}+\hat{\mathbf{g}}^{r}_{1,N;\sigma}\hat{\Omega}_{N,N}\left(\mathbb{1}-\left|t\right|^{2}\hat{\mathbf{g}}^{r}_{N,N;\sigma}\hat{\Omega}_{N,N}\right)^{-1}\left(\hat{\mathbf{g}}^{r}_{N,N;\sigma}+\left|t\right|^{2}\hat{\mathbf{g}}^{r}_{N,1;\sigma}\hat{\Omega}_{1,1}\hat{\mathbf{G}}^{r}_{1,N;\sigma}\right) \right]. \label{Eq:G1N_s}
\end{align}

Finally, we find $\hat{\mathbf{G}}^{r}_{1,1;\sigma}(\omega)$ and $\hat{\mathbf{G}}^{r}_{1,N;\sigma}(\omega)$ in Eq. (\ref{Eq:G11_s}) and Eq. (\ref{Eq:G1N_s}) in terms of finite matrix Greens functions that can be solved numerically

    \begin{align}
        \hat{\mathbf{G}}^{r}_{1,1;\sigma}(\omega)&=\left(\hat{\Lambda}_{1,1}-\left|t\right|^{4}\hat{\mathbf{g}}^{r}_{1,N;\sigma}\hat{\Omega}_{N,N}\hat{\Lambda}_{N;\sigma}^{-1}\hat{\mathbf{g}}^{r}_{N,1;\sigma}\hat{\Omega}_{1,1}\right)^{-1}\left(\hat{\mathbf{g}}^{r}_{1,1;\sigma}+\left|t\right|^{2}\hat{\mathbf{g}}^{r}_{1,N;\sigma}\hat{\Omega}_{N,N}\hat{\Lambda}_{N;\sigma}^{-1}\hat{\mathbf{g}}^{r}_{N,1;\sigma}\right), \label{Eq:G11_solved} \\ 
        \hat{\mathbf{G}}^{r}_{1,N;\sigma}(\omega)&=\left(\hat{\Lambda}_{1;\sigma}-\left|t\right|^{4}\hat{\mathbf{g}}^{r}_{1,N;\sigma}\hat{\Omega}_{N,N}\hat{\Lambda}_{N;\sigma}^{-1}\hat{\mathbf{g}}^{r}_{N,1;\sigma}\hat{\Omega}_{1,1}\right)^{-1}\left(\hat{\mathbf{g}}^{r}_{1,N;\sigma}+\left|t\right|^{2}\hat{\mathbf{g}}^{r}_{1,N;\sigma}\hat{\Omega}_{N,N}\hat{\Lambda}_{N;\sigma}^{-1}\hat{\mathbf{g}}^{r}_{N,N;\sigma}\right), \label{Eq:G1N_solved}
     \end{align}

where we have defined $\hat{\Lambda}_{j;\sigma} = \mathbb{1}-\left|t\right|^{2}\hat{\mathbf{g}}^{r}_{j,j,;\sigma}\hat{\Omega}_{j}$.

\section{Conductance}\label{Ap:Conductance}

In this Appendix, we present details of the calculation of the local and non-local conductances in  Eqs. (\ref{eq:conductance_GLL})-(\ref{eq:conductance_GRL}). For simplicity, in what follows we focus on the derivation of the current flowing through the left contact $I_L$ appearing  in Eq. (\ref{eq:current_G_lesserL}), the same procedure can be applied to the right lead. In addition, we  omit spin indices to simplify the notation. 

Using Dyson's equation in the Keldysh formalism using Eq. (\ref{eq:Hmix_leads}) as the perturbation connecting the (infinite) SE-SC-FMI system to the contact leads, the lesser Nambu  Green's functions  in Eq. (\ref{eq:current_G_lesserL}) can be expressed as
\begin{align}
    \hat{\mathbf{G}}_{L,1}^{<}&= t_L\left(\hat{\mathbf{g}}_{L,L}^{(0),t}\hat{\mathbf{G}}_{1,1}^{<}-\hat{\mathbf{g}}_{L,L}^{(0),<}\hat{\mathbf{G}}_{1,1}^{\bar{t}}\right),\label{eq:eq_DysonL1}\\ 
    \hat{\mathbf{G}}_{1,L}^{<}&= t_L\left(\hat{\mathbf{g}}_{L,L}^{(0),<}\hat{\mathbf{G}}_{1,1}^{t}-\hat{\mathbf{g}}_{L,L}^{(0),\bar{t}}\hat{\mathbf{G}}^<_{1,1}\right),\label{eq:eq_Dyson1L}
\end{align}
where the labels $t$ and $\bar{t}$ denote time-ordered and anti time-ordered Green's functions, respectively \cite{Haug_Jauho_book}. Substracting the left handsides, we obtain
\begin{align}
 \hat{\mathbf{G}}_{L,1}^{<}-  \hat{\mathbf{G}}_{1,L}^{<}&= t_L\left(\hat{\mathbf{g}}_{L,L}^{(0),t}+\hat{\mathbf{g}}_{L,L}^{(0),\bar{t}}\right)\hat{\mathbf{G}}_{1,1}^{<}-t_L \hat{\mathbf{g}}_{L,L}^{(0),<}\left(\hat{\mathbf{G}}_{1,1}^{t}+\hat{\mathbf{G}}_{1,1}^{\bar{t}}\right),\\
&= t_L\left(\hat{\mathbf{g}}_{L,L}^{(0),>}\hat{\mathbf{G}}_{1,1}^{<}-\hat{\mathbf{g}}_{L,L}^{(0),<}\hat{\mathbf{G}}_{1,1}^> \right),
\end{align}
where we have used the Keldysh Green's functions identity $\mathbf{G}^t+\mathbf{G}^{\bar{t}}=\mathbf{G}^>+\mathbf{G}^<$ \cite{Haug_Jauho_book}. Replacing this result in  Eq. (\ref{eq:current_G_lesserL}), and restoring the spin indices and the frequency arguments we obtain
\begin{align}
I_{L}&=-\frac{e}{h}t^2_{L}\sum_\sigma\int_{-\infty}^{\infty}d\omega\ \left[\hat{\mathbf{g}}_{L,L;\sigma}^{(0),<}(\omega)\hat{\mathbf{G}}_{1,1;\sigma}^{>}(\omega)-\hat{\mathbf{g}}_{L,L;\sigma}^{(0),>}(\omega)\hat{\mathbf{G}}_{1,1;\sigma}^{<}(\omega)\right]_{1,1},
\end{align}
which corresponds to Eq. (\ref{eq:current_G_lesserL2}). We recall that the notation $\left[\dots\right]_{1,1}$ denotes the first element of the Nambu  Green's function matrix. We now replace the expressions of the bare Green's functions $\hat{\mathbf{g}}_{L,L;\sigma}^{(0),<}(\omega)$ and $\hat{\mathbf{g}}_{L,L;\sigma}^{(0),>}(\omega)$ given by Eqs.  (\ref{eq:lesser_density}) and Eq. (\ref{eq:greater_density}) respectively, and obtain

    \begin{align}
        I_{L}&=-\frac{i e}{\pi \hbar}\gamma_{L}\sum_\sigma \int_{-\infty}^{\infty}d\omega\ \left[\begin{pmatrix}
        n_{L}(\omega) & 0\\
        0 & \bar{n}_{L}(\omega)
        \end{pmatrix}\begin{pmatrix}
        G_{1,1;\sigma}^{>}(\omega) & F_{1,1;\sigma}^{>}(\omega)\\
        \bar{F}_{1,1;\sigma}^{>}(\omega) & \bar{G}_{1,1;\sigma}^{>}(\omega)
        \end{pmatrix}+\right.\nonumber \\
&\quad  \quad \left. \begin{pmatrix}
        1-n_{L}(\omega) & 0\\
        0 & 1-\bar{n}_{L}(\omega)
        \end{pmatrix}\begin{pmatrix}
        G_{1,1;\sigma}^{<}(\omega) & F_{1,1;\sigma}^{<}(\omega)\\
        \bar{F}_{1,1;\sigma}^{<}(\omega) & \bar{G}_{1,1;\sigma}^{<}(\omega)\end{pmatrix}\right]_{1,1}, \\ 
        I_{L}&=-\frac{i e}{\pi \hbar}\gamma_{L}\int_{-\infty}^{\infty}d\omega\ \left[n_{L}(\omega)G_{1,1;\sigma }^{>}(\omega)+\left(1-n_{L}(\omega)\right)G_{1,1;\sigma}^{<}(\omega)\right] , \\ 
        I_{L}&=-\frac{i e}{\pi \hbar}\gamma_{L}\sum_\sigma \int_{-\infty}^{\infty}d\omega\ \left[n_{L}(\omega)\left(G_{1,1;\sigma}^{>}(\omega)-G_{1,1;\sigma}^{<}(\omega)\right)+G_{1,1;\sigma}^{<}(\omega)\right] , \\ 
        I_{L}&=-\frac{i e}{\pi \hbar}\gamma_{L}\sum_\sigma \int_{-\infty}^{\infty}d\omega\ \left[n_{L}(\omega)\left(G_{1,1;\sigma}^{r}(\omega)-G_{1,1;\sigma}^{a}(\omega)\right)+G_{1,1;\sigma}^{<}(\omega)\right]\label{eq:Iapendix} ,
    \end{align}

where we have used another Keldysh  identity $\mathbf{G}^{>}-\mathbf{G}^{<}=\mathbf{G}^{r}-\mathbf{G}^{a}$ \cite{Haug_Jauho_book}. In order to take the derivatives with respect to the chemical potentials  $\mu_{L}$ and $\mu_{R}$, the lesser Green's function in the right handside of Eq. (\ref{eq:Iapendix}) must be expanded in term of local lesser Green's function at the leads $\hat{\mathbf{g}}_{L,L}^{(0),<}(\omega)$ and $\hat{\mathbf{g}}_{R,R}^{(0),<}(\omega)$, for this purpose we use the  Dyson-Keldysh equation \cite{Cuevas96_Hamiltonian_approach_to_SC_contacts, Haug_Jauho_book}

\begin{align}
    \hat{\mathbf{G}}^{<}_{i,j;\sigma}(\omega)&=\left[\mathbb{1}+\hat{\mathbf{G}}^{r}_{i,k;\sigma}(\omega)\left(\hat{\mathcal{H}}^\text{w-leads}\right)_{k,\alpha}\right]\hat{\mathbf{g}}_{\alpha,\alpha;\sigma}^{(0), <}(\omega)\left[\mathbb{1}+\left(\hat{\mathcal{H}}^\text{w-leads}\right)_{\alpha,l} \hat{\mathbf{G}}^{a}_{l,j;\sigma}(\omega)\right],
\end{align}

where $\alpha=\{L,R\}$, and where $\hat{\mathcal{H}}^\text{w-leads}$ is the  BdG-Nambu Hamiltonian corresponding to Hamiltonian Eq. (\ref{eq:Hmix_leads}). Deriving the first element of the Nambu Green's function matrix in Eq. (\ref{eq:Iapendix}) allows to obtain 

\begin{align}
    \frac{d G_{1,1;\sigma}^{<}(\omega)}{dV_{L}}&=2 i \gamma_{L}\left(\frac{d n_{F}\left(\omega-eV_L\right)}{dV_{L}} G_{1,1;\sigma}^r(\omega)G_{1,1;\sigma}^a(\omega) + \frac{d\bar{n}_{F}\left(\omega-eV_L\right)}{dV_{L}}F_{1,1;\sigma}^r(\omega)\bar{F}_{1,1;\sigma}^a(\omega) \right),\label{eq:derivativeL}\\ 
    \frac{d G_{1,1;\sigma}^{<}(\omega)}{dV_{R}}&=2 i \gamma_{R} \left(\frac{d n_{F}\left(\omega-eV_R\right)}{dV_{R}} G_{1,N;\sigma}^r(\omega)G_{N,1;\sigma}^a(\omega) + \frac{d\bar{n}_{F}\left(\omega-eV_R\right)}{dV_{R}}F_{1,N;\sigma}^r(\omega)\bar{F}_{N,1;\sigma}^a(\omega) \right).\label{eq:derivativeR}
\end{align}    

Finally, replacing Eq. (\ref{eq:derivativeL}) and Eq. (\ref{eq:derivativeR}) into Eq. (\ref{eq:Iapendix}), the local and non-local conductance depend explicitly on the derivatives of the Fermi distribution 

\begin{align}
\mathsf{G}_{LL}&=\frac{2e \gamma_L}{\pi \hbar}\sum_\sigma \int_{-\infty}^{\infty}d\omega \left\{-\pi\rho_{1,1;\sigma}(\omega)\frac{d n_{F}\left(\omega-eV_L\right)}{dV_{L}}\right.\nonumber\\
&\left.+\gamma_{L}\left(\frac{d n_{F}\left(\omega-eV_L\right)}{dV_{L}} G_{1,1;\sigma}^r(\omega)G_{1,1;\sigma}^a(\omega) + \frac{d\bar{n}_{L}\left(\omega-eV_L\right)}{dV_{L}}F_{1,1;\sigma}^r(\omega)\bar{F}_{1,1;\sigma}^a(\omega) \right)\right\}\label{eq:G_LL_final3},\\
\mathsf{G}_{LR}&=\frac{2e \gamma_L \gamma_R}{\pi \hbar}\sum_\sigma \int_{-\infty}^{\infty}d\omega \ \left(\frac{d n_{F}\left(\omega-eV_R\right)}{dV_{R}} G_{1,N;\sigma}^r(\omega)G_{N,1;\sigma}^a(\omega) + \frac{d\bar{n}_{F}\left(\omega-eV_R\right)}{dV_{R}}F_{1,N;\sigma}^r(\omega)\bar{F}_{N,1;\sigma}^a(\omega) \right)\label{eq:G_LR_final3},
\end{align} 

Taking the limit $T\rightarrow0$, the derivatives of the Fermi function become $dn_F(\omega-eV_\alpha)/dV_\alpha\rightarrow e\delta(\omega-eV_\alpha)$,  allowing a straightforward integration. Therefore,  the Green's functions that appear in the expressions of conductances Eqs. (\ref{eq:conductance_GLL}) and  (\ref{eq:conductance_GLR}) are evaluated in terms of the gates potentials $V_L$ and $V_R$. 

\section{Finite-temperature effects}\label{sec:finiteT}

In order to quantify the effect of a finite temperature, we have explicitly calculated the local conductance $\mathsf{G}_{LL}$ in Eq. (\ref{eq:G_LL_final3}) for a system with $N_\text{M}=38$, for two different parameters $\mu_\text{BG}/\Delta=-2$ and $\mu_\text{BG}/\Delta=-8$, and for different temperatures indicated in the insets (the results for the non-local conductance $\mathsf{G}_{LR}$ are qualitatively similar and not more illuminating). 
The derivative of the Fermi function appearing in Eqs. (\ref{eq:G_LL_final3}) and (\ref{eq:G_LR_final3}), which at $T=0$ reduces to a Dirac delta function centered at $\omega=eV_L$, becomes broadened at finite temperature, with a characteristic width of order $k_B T$. Consequently, the integral effectively turns into a convolution with a thermal ``smearing'' function $d n_F(\omega-V_L) / dV_L$. Our calculations indeed show that increasing the temperature smears the conductance peaks, reducing their intensity and making the particle-hole peaks less distinguishable (note that a non-vanishing residual width at $T=0$ persists in these plots, due to the effect of a finite broadening $\gamma_L$ and $\gamma_R$). This, in turn, makes the identification of the zero-crossing point associated with the phase transition more challenging. However, we stress that the position of the zero-crossing point does not change in relation to the parameter $\mu_\text{BG}/\Delta$, and therefore our main conclusions in Fig. \ref{fig:conductance for h=1.5 4L} are not qualitatively affected.

\begin{figure}[h!]
    \centering
        \includegraphics[width=0.65\textwidth]{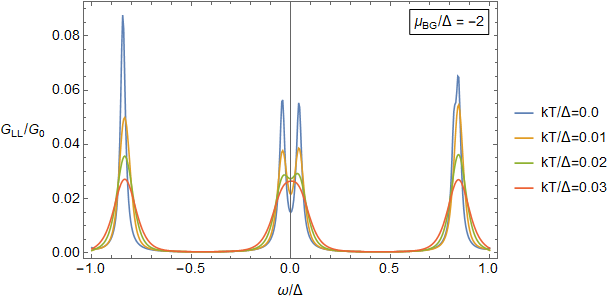} 
        \includegraphics[width=0.65\textwidth]{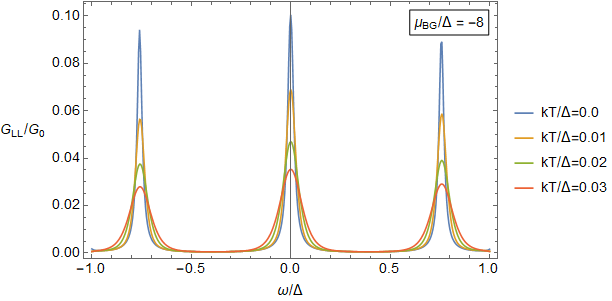}
\caption{(color online). Local conductance $\mathsf{G}_{LL}$ as a function of $\omega/\Delta$ for a system with $N_\text{M}=38$, computed for two different parameters $\mu_\text{BG}/\Delta=-2$ and $\mu_\text{BG}/\Delta=-8$, and different temperatures, as given by Eq. (\ref{eq:G_LL_final3}). \label{fig:thermal_smearing}}
\end{figure}

\end{widetext}

%

\end{document}